\documentclass[10pt]{article}%
\usepackage[english]{babel}
\usepackage{times}
\usepackage{pifont}
\usepackage{cite}
\usepackage[usenames]{color}
\usepackage{bbm,amsmath,amssymb,amsthm,amsbsy}
\usepackage{graphicx}
\usepackage[caption=false,font=small,labelfont=bf,textfont=sf]{subfig}
\usepackage[colorlinks = true]{hyperref}

\usepackage[toc,page,header]{appendix}
\usepackage{fullpage}
\numberwithin{equation}{section}

\newtheorem{proposition}{Proposition}
\newtheorem{lemma}{Lemma}
\newtheorem{claim}{Claim}
\newtheorem{theorem}{Theorem}
\newtheorem{corollary}{Corollary}

\newtheorem{definition}{Definition}
\theoremstyle{remark}

\newcommand{\tmop}[1]{\ensuremath{\operatorname{#1}}}
\newcommand  \stack[2]  {\overset{\text{#1}}{#2}}
\DeclareMathOperator*{\minimize}{\operatorname{minimize}}

\DeclareMathOperator*{\argmin}{\operatorname{arg~min}}
\DeclareMathOperator*{\argmax}{\operatorname{arg~max}}
\def \st {\operatorname*{subject\ to\ }}
\def \nn        {\nonumber}

\def \R {{\mathbb{R}}}

\def \A {{\mathcal{A}}}
\def \O {{\mathcal{O}}}

\def \inv       {^{\text{-1}}}
\def \eye       {\mathbf{I}}
\def \zero      {\mathbf{0}}

\def \lg        {\langle}
\def \rg        {\rangle}
\def \diag      {\tmop{diag}}
\def \tr        {\tmop{trace}}
\def \rank      {\tmop{rank}}

\def \range     {\tmop{Range}}

\def \vlambda   {{\boldsymbol{\lambda}}}

\def \mPhi      {{\boldsymbol{\Phi}}}
\def \mPsi      {{\boldsymbol{\Psi}}}
\def \mSigma    {{\boldsymbol{\Sigma}}}

\newcommand{\vct}[1]{\mathbf{#1}}

\def \p {\vct{p}}
\def \q {\vct{q}}

\def \s {\vct{s}}
\def \y {\vct{t}}

\def \v {\vct{v}}

\def \x {\vct{x}}
\def \y {\vct{y}}

\newcommand{\mtx}[1]{\mathbf{#1}}
\def \mA {\mtx{A}}
\def \mB {\mtx{B}}

\def \mD {\mtx{D}}
\def \mE {\mtx{E}}

\def \mG {\mtx{G}}
\def \mH {\mtx{H}}

\def \mP {\mtx{P}}
\def \mQ {\mtx{Q}}
\def \mR {\mtx{R}}
\def \mS {\mtx{S}}

\def \mU {\mtx{U}}
\def \mV {\mtx{V}}
\def \mW {\mtx{W}}
\def \mX {\mtx{X}}
\def \mY {\mtx{Y}}
\def \mZ {\mtx{Z}}

\newcommand{\set}[1]{{\mathcal{#1}}}

\def \calE {\set{E}}

\def \calQ {\set{Q}}

\def \calX {\set{X}}

\def \Poff {\mathcal{P_{\tmop{off}}}}
\def \Pon  {\mathcal{P_{\tmop{on}}}}
\def \dist {\tmop{d}}
\def \wh   {\widehat}

\begin{document}
\title{The Non-convex Geometry of Low-rank Matrix Optimization}
\author{{\sc Qiuwei Li} \qquad {\sc Zhihui Zhu}    \qquad {\sc Gongguo Tang}\thanks{This work was supported by the National Science Foundation [CCF-1704204 to G.T. and Q.L., CCF-1409261 to Z.Z.].}
\\[2pt]
{Department of Electrical Engineering, Colorado School of Mines, CO, USA}
}
\date{}
\maketitle

\begin{abstract}
This work considers two popular minimization problems: (i) the minimization of a general convex function $f(\mX)$ with the domain being positive semi-definite matrices; (ii) the minimization of a general convex function $f(\mX)$ regularized by the matrix nuclear norm $\|\mX\|_*$ with the domain being general matrices. Despite their optimal statistical performance in the literature, these two optimization problems have a high computational complexity  even when solved  using tailored fast convex solvers. To develop faster and more scalable algorithms, we follow the proposal of Burer and Monteiro to factor the low-rank variable $\mX = \mU\mU^\top $ (for semi-definite matrices) or $\mX=\mU\mV^\top $ (for general matrices) and also replace the nuclear norm $\|\mX\|_*$ with $(\|\mU\|_F^2+\|\mV\|_F^2)/2$. In spite of the non-convexity of the resulting factored formulations, we prove that each critical point either corresponds to the global optimum of the original convex problems  or is a strict saddle where the Hessian matrix  has a strictly negative eigenvalue. Such a nice geometric structure of the factored formulations allows many local search algorithms to find a global optimizer even with random initializations.


\end{abstract}

\section{Introduction}

Nonconvex reformulations of convex optimization problems have received a surge of renewed interest   for efficiency and scalability reasons~\cite{nvx:sun2016nonconvex, nvx:sun2016geometric,nvx:sun2015complete,nvx:ge2016matrix,nvx:ge2017no,nvx:tu2015low,nvx:bhojanapalli2015dropping,nvx:li2016symmetry,nvx:li2016overcomplete,nvx:kyrillidis2017provable,nvx:park2016finding,nvx:park2016provable,nvx:zhao2015nonconvex,nvx:wang2017unified,nvx:de2015global,nvx:tran2016extended,nvx:Li2017tensor}.  Compared
with the convex formulations, the non-convex ones typically involve many fewer variables, allowing them to scale to scenarios with millions of variables. Besides, simple algorithms~\cite{nvx:lee2016gradient,nvx:ge2015escaping, nvx:sun2016nonconvex} applied to the non-convex formulations have surprisingly good performance in practice. However,
a complete understanding of this phenomenon, particularly the geometrical structures of these non-convex optimization problems, is still an active research area. Unlike the simple geometry of convex
optimization problems where local minimizers are also global ones, the landscapes of general non-convex functions can become extremely complicated. Fortunately, for a range of convex optimization
problems, particularly for matrix completion and sensing problems, the corresponding non-convex
reformulations have nice geometric structures that allow local-search algorithms to converge to global
optimality~\cite{nvx:lee2016gradient,nvx:ge2015escaping, nvx:sun2016nonconvex,nvx:ge2016matrix,nvx:ge2017no,nvx:li2016symmetry,nvx:zhu2017global} .

We extend this line of investigation by working with a general convex function $f(\mX)$ and considering the following two popular optimization problems:
\begin{align}
\text{ For symmetric case: }\minimize_{\mX\in\R^{n\times n}}&\ f(\mX)\ \st \mX\succeq 0\tag{$\mathcal{P}_0$}\label{nvx:eqn:P0}\\
\text{ For nonsymmetric case: }\minimize_{\mX\in\R^{n\times m}}&\ f(\mX) + \lambda\|\mX\|_*  ~\text{ where }\lambda>0 \tag{$\mathcal{P}_1$} \label{nvx:eqn:P1}
\end{align}
For these two problems, even fast first-order methods, such as the projected gradient descent  algorithm~\cite{nvx:boyd2004convex}, require performing an expensive eigenvalue decomposition or singular value decomposition in each iteration. These expensive operations form the major computational bottleneck and prevent them from scaling to scenarios with  millions of variables,  a typical situation in a diverse range  of applications, including quantum state tomography~\cite{nvx:gross2010quantum}, user preferences prediction ~\cite{nvx:decoste2006collaborative}, and pairwise distances estimation in sensor localization~\cite{nvx:biswas2004semidefinite}.

\subsection{Our Approach: Burer-Monteiro Style Parameterization}
As we have seen, the extremely large dimension of the optimization variable $\mX$ and the accordingly
expensive eigenvalue or singular value decompositions on X form the major computational bottleneck
of the convex optimization algorithms. An immediate question might be ``Is there a way to directly 
reduce the dimension of the optimization variable $\mX$ and meanwhile avoid performing the expensive
eigenvalue or singular value decompositions?"

This question can be answered when the original optimization problems~\eqref{nvx:eqn:P0}-\eqref{nvx:eqn:P1} admit a low-rank solution $\mX^\star$ with $\rank(\mX^\star)=r^\star\ll \min\{n,m\}$. Then we can follow the proposal of Burer and Monteiro~\cite{nvx:burer2003nonlinear} to parameterize the low-rank variable as $\mX = \mU\mU^\top $ for~\eqref{nvx:eqn:P0} or $\mX=\mU\mV^\top $ for~\eqref{nvx:eqn:P1},
where $\mU \in \R^{n\times r}$ and $\mV \in\R^{m\times r}$ with $r\geq r^\star$.
Moreover, since $\|\mX\|_*=\minimize_{\mX=\mU\mV^\top }(\|\mU\|_F^2+\|\mV\|_F^2)/2$, we obtain the following non-convex re-parameterizations of~\eqref{nvx:eqn:P0}-\eqref{nvx:eqn:P1}:
\begin{align}
\text{ For  symmetric case: } \quad&\minimize_{\mU \in \R^{n\times r}} g(\mU)=f(\mU\mU^\top ) \tag{$\mathcal{F}_0$}\label{nvx:eqn:F0}\\
\text{ For nonsymmetric case: } \quad& \minimize_{\mU \in \R^{n\times r},V\in\R^{m\times r}} g(\mU,\mV)=f(\mU\mV^\top )+ \frac{\lambda}{2}\left(\|\mU\|_F^2+\|\mV\|_F^2\right)\tag{$\mathcal{F}_1$}\label{nvx:eqn:F1}
\end{align}
Since $r\ll \{p,q\}$, the resulting factored problems~\eqref{nvx:eqn:F0}-\eqref{nvx:eqn:F1} involve many fewer variables. Moreover, because the positive semi-definite constraint is removed from~\eqref{nvx:eqn:P0} and the nuclear norm $\|\mX\|_*$ in~\eqref{nvx:eqn:P1} is replaced by $(\|\mU\|_F^2+\|\mV\|_F^2)/2$, there is no need to perform an eigenvalue (or a singular value) decomposition in solving the factored problems.

The past two years have seen renewed interest in the Burer-Monteiro factorization for solving low-rank matrix optimization problems~\cite{nvx:ge2016matrix,nvx:ge2017no,nvx:bhojanapalli2015dropping,nvx:tu2015low,nvx:li2016symmetry,nvx:li2017low}. With technical innovations in analyzing the non-convex landscape of the factored objective function, several recent works have shown that with an exact parameterization (i.e., $r = r^\star$)  the resulting factored reformulation has no spurious local minima or degenerate saddle points~\cite{nvx:ge2016matrix,nvx:ge2017no,nvx:li2016symmetry,nvx:zhu2017global}. An important implication is that local-search algorithms such as gradient descent and its variants can converge to the global optima with even random initialization~\cite{nvx:lee2016gradient,nvx:ge2015escaping, nvx:sun2016nonconvex}.

We generalize this line of work by assuming a general objective function $f(\mX)$ in~\eqref{nvx:eqn:P0}-\eqref{nvx:eqn:P1}, not necessarily coming from a matrix inverse problem. This generality allows us to view the resulting factored problems~\eqref{nvx:eqn:F0}-\eqref{nvx:eqn:F1} as a way to solve the original convex optimization problems to the global optimum, rather than a new modeling method. This perspective, also taken by Burer and Monteiro in their original work~\cite{nvx:burer2003nonlinear}, frees us from rederiving the statistical performances of the resulting factored optimization problems. Instead, the statistical performances of the resulting factored optimization problems inherit from that of the original convex optimization problems, whose statistical performance can be analyzed using a suite of powerful convex analysis techniques, which have accumulated from several decades of research.
For example, the original convex optimization problems~\eqref{nvx:eqn:P0}-\eqref{nvx:eqn:P1} have information-theoretically optimal sampling complexity~\cite{nvx:candes2010power}, achieve minimax denoising rate~\cite{nvx:candes2010matrix} and satisfy tight oracle inequalities~\cite{nvx:candes2011tight}.
Therefore, the statistical performances of  the factored optimization problems~\eqref{nvx:eqn:F0}-\eqref{nvx:eqn:F1} share the same theoretical bounds as those of the original convex optimization problems~\eqref{nvx:eqn:P0}-\eqref{nvx:eqn:P1}, as long as we can show that the two problems are equivalent.

In spite of their optimal statistical performance~\cite{nvx:davenport2016overview,nvx:candes2010matrix,nvx:candes2010power,nvx:candes2011tight}, the original convex optimization problems cannot be scaled to solve the practical problems that originally motivate their development even with specialized first-order algorithms. This was realized since the advent of this field where the low-rank factorization method was proposed as an alternative to convex solvers~\cite{nvx:burer2003nonlinear}. When coupled with stochastic gradient descent, low-rank factorization leads to state-of-the-art performance in practical matrix recovery problems~\cite{nvx:ge2016matrix,nvx:ge2017no,nvx:li2016symmetry,nvx:zhu2017global,nvx:tu2015low}. Therefore, our general analysis technique also sheds light on the connection between the geometries of the original convex programs and their non-convex reformulations.

Although the Burer-Monteiro  parameterization tremendously reduces the number of optimization variables from $n^2$ to $nr$ (or $nm$ to $(n+m)r$) when $r$ is very small, the intrinsic bi-linearity
makes the factored objective functions non-convex and introduces additional critical points that are not global optima of the factored optimization problems. One of our main purposes is to show that these additional critical points will not introduce spurious local minima. More precisely, we want to figure out what properties of the convex function $f$ are required for the factored objective functions $g$ to have no spurious local minima.

	\subsection{Enlightening Examples}\label{nvx:sec:enlightening examples}
	To gain some intuition about the properties of  $f$ such that the factored objective function $g$ has no spurious local minima (which is one of the main goals considered in this paper), let us consider the following two examples:  Weighted principal component analysis (weighted PCA) and the matrix sensing problem.
	\paragraph{Weighted PCA:} Consider the symmetric weighted PCA problem in which the lifted objective function is
	\[f(\mX)=\frac{1}{2}\|\mW\odot(\mX-\mX^\star)\|_F^2,\]
	where $\odot$ is the Hadamard product,  $\mX^\star$ is the global optimum we want to recover and $\mW$ is the known weighting matrix (which is assumed to have no zero entries for simplicity).
	After applying the Burer-Monteiro  parameterization to  $f(\mX)$, we obtain the factored objective function
	\begin{align*}
	g(\mU)=\frac{1}{2}\|\mW\odot(\mU\mU^\top-\mX^\star)\|_F^2.
	\end{align*}
	To investigate the conditions under which the  bi-linearity $\phi(\mU)=\mU\mU^\top$  will (not) introduce additional local minima to the factored optimization problems, consider a simple (but enlightening) two-dimensional example where
	$
	\mW=\begin{bmatrix}\sqrt{1+a}&1\\1&\sqrt{1+a}\end{bmatrix} \text{ for some }a\geq0,     \mX^\star=\begin{bmatrix}1&1\\1&1\end{bmatrix},$ and     $\mU=\begin{bmatrix}x\\y\end{bmatrix}$ for unknowns $x,y$.
	Then the factored objective function becomes
	\begin{align}
	\label{nvx:eqn:pca:g:2}
	g(\mU)&
	=\frac{1+a}{2}\left(x^2-1\right)^2+\frac{1+a}{2}\left(y^2-1\right)^2+  (x y-1)^2.
	\end{align}
	In this particular setting, we will see that  the value of $a$ in the weighting matrix is the deciding factor for the occurrence of spurious local minima.
	\begin{claim}\label{nvx:claim}
		The factored objective function $g(\mU)$ in~\eqref{nvx:eqn:pca:g:2}
		has no spurious local minima when $a\in[0,2)$; while for $a>2$, spurious local minima will appear.
	\end{claim}
	\begin{proof}
		First of all, we compute the gradient $\nabla g(\mU)$ and Hessian $\nabla^2 g(\mU)$:
		\begin{align*}
		\nabla g(\mU)&
		=2\begin{bmatrix}
		(a+1) \left(x^2-1\right) x+  y (x y-1)\\
		(a+1) \left(y^2-1\right) y+  x (x y-1)
		\end{bmatrix},
		\\
		\nabla^2g(\mU)&=
		2\begin{bmatrix}
		y^2+\left(3 x^2-1\right) (a+1) & 2 x y-1 \\
		2 x y-1 &   x^2+\left(3 y^2-1\right) (a+1)\\
		\end{bmatrix}.
		\end{align*}
		Now we  collect all the critical points by solving $\nabla g(\mU)=0$ and list the Hessian of $g$ at these points as follows\footnote{Note that if $\mU$ is a critical point, so is $-\mU$, since $\nabla g(-\mU)=-\nabla g(\mU)$. Hence we only list one part of these critical points.}
		\begin{itemize}
			\item[\ding{172}] $\mU_1=(0,0)$,    $\nabla^2g(\mU_1)=-2\begin{bmatrix}
			a+1 & 1 \\
			1  &   a+1\\
			\end{bmatrix};$
			\item[\ding{173}] $\mU_2=(1,1)$,     $\nabla^2g(\mU_2)=2
			\begin{bmatrix}
			2 a+3 & 1 \\
			1 & 2 a+3 \\
			\end{bmatrix};$
			\item[\ding{174}] $\mU_3=(\sqrt{\frac{a}{a+2}},  -\sqrt{\frac{a}{a+2}})$,   $\nabla^2 g(\mU_3)=
			\begin{bmatrix}
			4 a+\frac{8}{a+2}-6 & \frac{8}{a+2}-6 \\
			\frac{8}{a+2}-6 & 4 a+\frac{8}{a+2}-6 \\
			\end{bmatrix};$
			\item[\ding{175}] $\mU_4=(\frac{\sqrt{\frac{\sqrt{a^2-4}+a}{a}}}{\sqrt{2}},  -\frac{\sqrt{2}}{a \sqrt{\frac{\sqrt{a^2-4}+a}{a}}})$,
			$\nabla^2 g(\mU_4)=
			\begin{bmatrix}
			a+3 \sqrt{a^2-4}+2+\frac{2 \sqrt{a^2-4}}{a} & -\frac{2 (a+2)}{a} \\
			-\frac{2 (a+2)}{a} & a-3 \sqrt{a^2-4}+2-\frac{2 \sqrt{a^2-4}}{a} \\
			\end{bmatrix}.$
		\end{itemize}
		Note that the critical point $\mU_4$ exists only for $a\geq 2$.
		By checking the signs of the two eigenvalues (denoted by $\lambda_1$ and $\lambda_2$) of these Hessians, we can further classify these critical points as a local minimum,  a local maximum, or a saddle point\footnote{This classification of the critical points using the Hessian information is known as the second derivative test, which says a critical point is a local maximum if the Hessian is negative definite, a local minimum is the Hessian is positive definite, and a saddle point if the Hessian matrix has both positive and negative eigenvalues.}:
		\begin{itemize}
			\item[\ding{172}]   $\lambda_1=-2(a+2),\lambda_2=-2a$. So, $\mU_1$ is a local maximum for $a>0$ and a strict saddle for $a=0$ (see Definition \ref{nvx:def:strict}).
			\item[\ding{173}] $\lambda_1=4 (a+1)>0,\lambda_2=4 (a+2)>0.$ So, $\mU_2$ is a local minimum (also a global minimum as $g(\mU_2)=0$).
			\item[\ding{174}]  $\lambda_1=\frac{4 (a-2) (a+1)}{a+2}\begin{cases}<0, & a\in[0,2)\\ >0, &a>2 \end{cases},\lambda_2=4 a>0$. So,
			$\mU_3$ is
			$\begin{cases}\text{a saddle point}, & a\in[0,2)\\\text{a \emph{spurious} local minimum}, &a>2 \end{cases}$
			\item[\ding{175}] From the determinant, we have
			$\lambda_1\cdot\lambda_2=-\frac{8 (a-2) (a+1) (a+2)}{a}<0$ for $a>2$. So, $\mU_4$ is a saddle point for $a>2$.
		\end{itemize}
	\end{proof}

\begin{figure}[ht]
		\begin{center}
			\subfloat[Small dynamic range]{\includegraphics[width=0.23\textwidth,  clip = true]{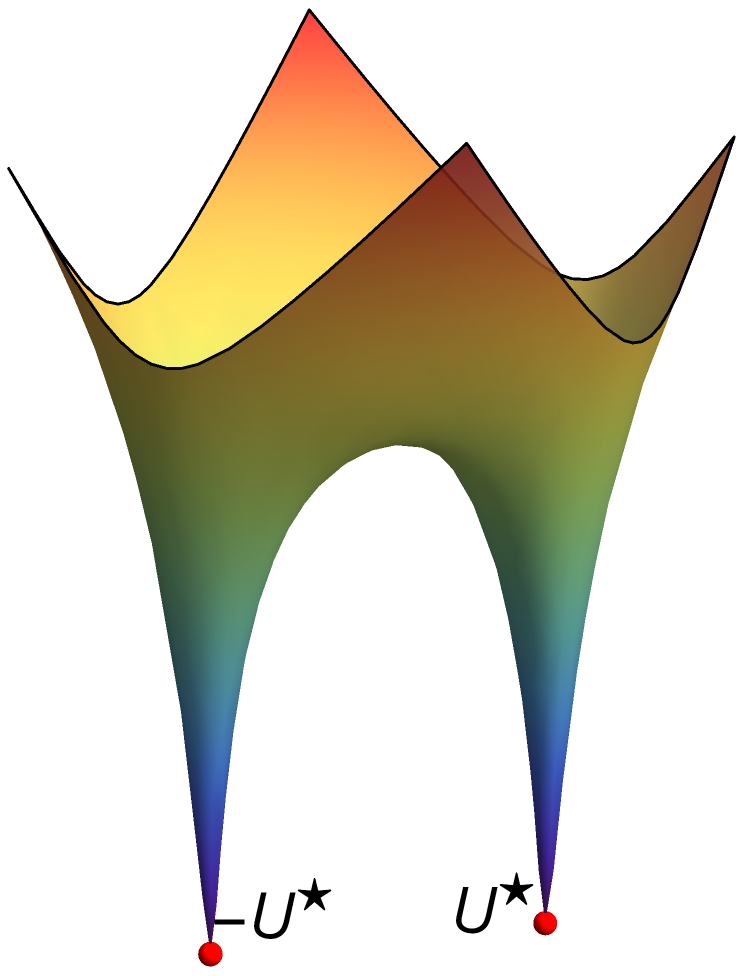}
			}
			\qquad\qquad
			\subfloat[Large dynamic range]{\includegraphics[width=0.23\textwidth,  clip = true]{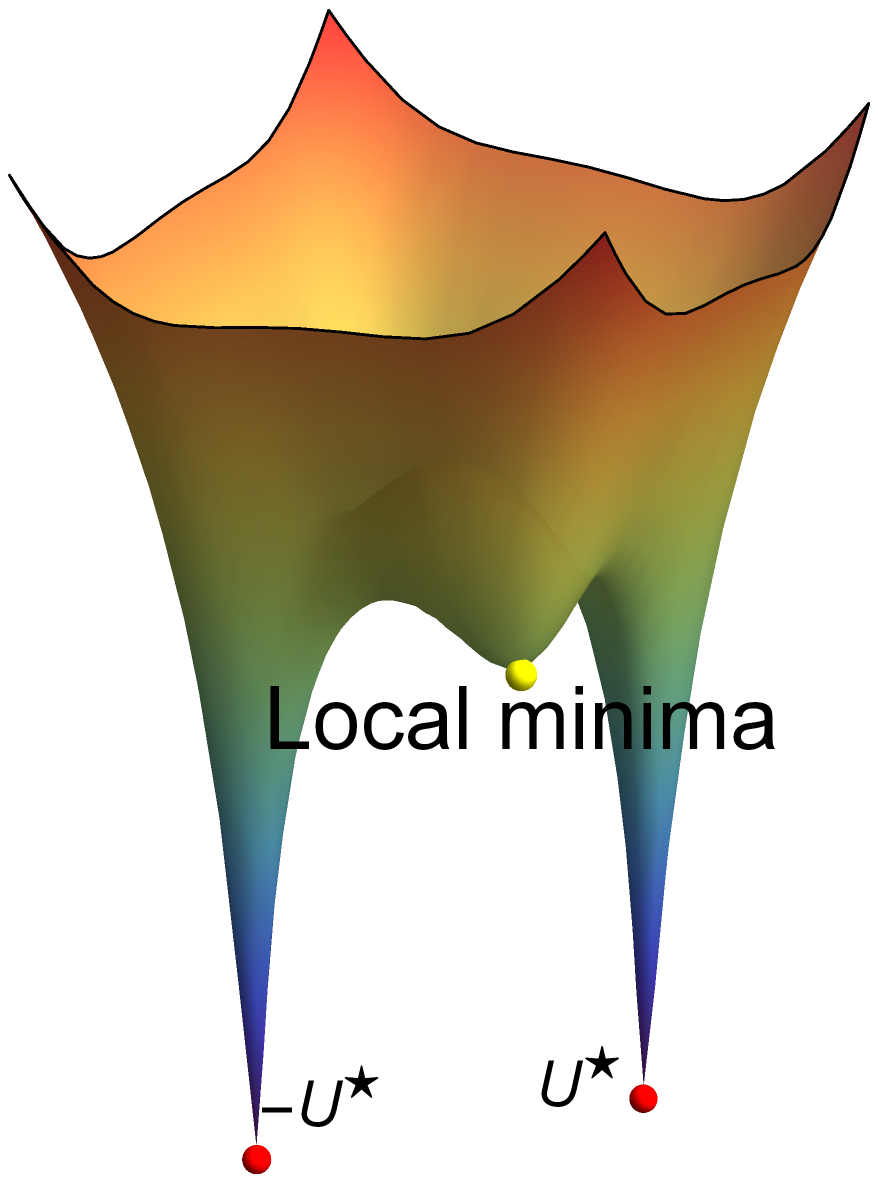}
			}
			\caption[Foo content]{Factored function landscapes corresponding to different  dynamic ranges of the weights $\mW$: (a) a small dynamic range with  $ {\max W_{ij}^2}/{\min W_{ij}^2}=1$ and (b) a large dynamic range with  $ {\max W_{ij}^2}/{\min W_{ij}^2}>3$.}
			\label{nvx:fig:local:and:nolocal}
		\end{center}
	\end{figure}

	In this example, the value of $a$ controls the dynamic range of the weights as  $ {\max W_{ij}^2}/{\min W_{ij}^2}=1+a$. Therefore, Claim \ref{nvx:claim} can be interpreted as a relationship between the spurious local minima and the dynamic range:
	if the dynamic range   ${\max W_{ij}^2}/{\min W_{ij}^2}$ is smaller than 3, there will be no spurious local minima; while if the dynamic range is larger than 3,  spurious local minima will appear.
	We also plot the landscapes of the factored objective function $g(\mU)$ in~\eqref{nvx:eqn:pca:g:2} with different dynamic ranges in Figure \ref{nvx:fig:local:and:nolocal}.

	As we have seen, the dynamic range of the weighting matrix serves as a determinant factor for the appearance of the spurious local minima for $g(\mU)$ in~\eqref{nvx:eqn:pca:g:2}.  To extend the above observations   to general objective functions, we now  interpret this condition (on the dynamic range of the weighting matrix) by relating it with the condition number of the Hessian matrix $\nabla^2 f(\mX)$. This can be seen from
	the following directional-curvature form for $f(\mX)$
	\[[\nabla^2 f(\mX)](\mD,\mD)=\|\mW\odot \mD\|_F^2,\]
	where $[\nabla^2 f(\mX)](\mD,\mD)$ is the directional curvature of $f(\mX)$ along the matrix $\mD$ of the same dimension as $\mX$, defined by $\sum_{i,j,l,k}\frac{\partial^2 f(\mX)}{\partial X_{ij}\partial X_{lk}} D_{ij}D_{lk}.$
	This implies that the condition number $\lambda_{\max}(\nabla^2 f(\mX))/\lambda_{\min}(\nabla^2 f(\mX))$  is upper-bounded by this dynamic range:
	\begin{align}\label{nvx:eqn:wpca:q}
	\min_{ij}|W_{ij}|^2\cdot\|\mD\|_F^2 \leq[\nabla^2 f(\mX)](\mD,\mD)\leq \max_{ij}|W_{ij}|^2\cdot\|\mD\|_F^2\quad\Leftrightarrow \quad \frac{\lambda_{\max}(\nabla^2 f(\mX))}{\lambda_{\min}(\nabla^2 f(\mX))}\leq \frac{\max W_{ij}^2}{\min W_{ij}^2}
	\end{align}
	Therefore,  we conjecture that the condition number of the general convex function $f(\mX)$ would be a deciding factor of the behavior of the landscape of the factored objective function and a large condition number is very likely to introduce spurious local minima to the factored problem.

\paragraph{Matrix Sensing:} The above conjecture can be further verified by the matrix sensing problem where the goal is to recover the low rank PSD matrix $\mX^\star\in\R^{n\times n}$ from the linear measurement $\y=\A(\mX^\star)$ with $\A: \R^{n\times n}\to\R^m$ being a linear measurement operator. Consider the factored objective function  $g(\mU)=f(\mU\mU^\top )$ with $U\in\R^{n\times r}$. In~\cite{nvx:li2016symmetry,nvx:bhojanapalli2016lowrankrecoveryl}, the authors showed that the non-convex parametrization $\mU\mU^\top$ will not introduce spurious local minima to the factored objective function, provided  the linear measurement operator $\A$ satisfies the following restricted isometry property (RIP).
\begin{definition}[Restricted Isometry Property]
	A linear operator $\A: \R^{n\times n}\to\R^m$ satisfies the $r$-RIP with constant $\delta_r$ if
	\begin{align}\label{nvx:RIP}
	(1-\delta_r)\|\mD\|_F^2\leq \|\A(\mD)\|_2^2\leq (1+\delta_r)\|\mD\|_F^2
	\end{align}
	holds for all $n\times n$ matrices $\mD$ with $\rank(\mD)\leq r$.
\end{definition}
Note that the required condition~\eqref{nvx:RIP}  essentially says that the  condition number of Hessian matrix $\nabla^2 f(\mX)$ should be small at least in the directions of the low-rank matrices $\mD$, since the directional curvature form of $f(\mX)$ is computed as $[\nabla^2 f(\mX)](\mD,\mD)=\|\A(\mD)\|_F^2$.

From these two examples, we see that as long as the Hessian matrix of the original convex function $f(\mX)$ has a small (restricted) condition number, the resulting factored objective function has a landscape such that all local minima correspond to the globally optimal solution. Therefore, we believe that such a restricted well-conditionedness property might be the key factor bring us a benign factored landscape, i.e.,
\begin{align*}
&\alpha\|\mD\|_F^2\leq [\nabla^2 f(\mX)](\mD,\mD)\leq \beta \|\mD\|_F^2 \text{ with some small } \beta/\alpha,
\end{align*}
which says that the landscape of $f(\mX)$ in the lifted space is bowl-shaped, at least in the directions of low-rank matrices.

\subsection{Our Results}

Before presenting the main results, we list a few necessary definitions.
\begin{definition}[\bf Critical points] For a continuous function $f:\R^n\to\R$,  we say $\x\in\R^n$ is a critical point of function $f$, if the gradient vanishes, i.e., $\nabla f(\x)=\zero$.
\end{definition}
\begin{definition}[\bf Strict saddles; or ridable saddles~\cite{nvx:sun2016nonconvex}]
	For a twice differentiable function $f$, a critical point $\x$ is a strict saddle if the Hessian matrix $\nabla^2 f(\x)$ has at least one strictly negative eigenvalue.
	\label{nvx:def:strict}
\end{definition}

\begin{definition}[\bf Strict saddle property ~\cite{nvx:ge2016matrix}]\label{nvx:def:strict:saddle:property}
	A twice differentiable function satisfies strict saddle property if each critical point either corresponds to the local minima or is a strict saddle.
\end{definition}

Heuristically, the strict saddle property describes a geometric structure of the landscape: if a critical point is not a local minimum, then it is a strict saddle, which implies that the Hessian matrix at this point has a strictly negative eigenvalue. Hence, we can continue to decrease the function value at this point along the negative-curvature direction.
This nice geometric structure  ensures that many local-search algorithms, such as noisy gradient descent~\cite{nvx:ge2015escaping}, vanilla gradient descent with random initialization~\cite{nvx:lee2016gradient}  and the trust region method~\cite{nvx:sun2016nonconvex},
can escape from all the saddle points along the directions associated with the Hessian's negative eigenvalues, and hence converge to a local minimum.
\begin{theorem}[\bf Local convergence for strict saddle property ~\cite{nvx:ge2015escaping,nvx:sun2016nonconvex,nvx:lee2016gradient,nvx:jin2017escape,nvx:lee2017first}]\label{nvx:thm:informal}
	The strict saddle property\footnote{To be precise, Lee et al.~\cite{nvx:lee2017first} showed that for any function that has a Lipschitz continuous gradient and obeys the strict saddle property, first-order methods with a random initialization almost always escape all the saddle points and converge to a local minimum. The   Lipschitz-gradient assumption is commonly adopted for analyzing the convergence of local-search algorithms, and we will discuss this issue after Theorem \ref{nvx:thm:main:1}. To obtain explicit convergence rate, other properties (like the gradient at the points that are away from the critical points is not small) about the objective functions may be required~\cite{nvx:ge2015escaping,nvx:sun2016nonconvex,nvx:jin2017escape,nvx:du2017gradient}. In this paper, similar to~\cite{nvx:ge2016matrix}, we mostly focus on the properties of the critical points, and we omit the details about the convergence rate. However, we should note that, by utilizing the similar approach in~\cite{nvx:zhu2017global}, it is possible to extend the strict saddle property so that we can obtain explicit convergence rate for certain algorithms~\cite{nvx:ge2015escaping,nvx:sun2016nonconvex,nvx:jin2017escape} when applied for solving the factored low-rank problems.} allows many local-search algorithms to escape all the saddle points and converge to a local minimum.
\end{theorem}

Our primary interest is to understand how the original convex landscapes are transformed by the factored parameterization $\mX = \mU\mU^\top $ or $\mX=\mU\mV^\top $, particularly how the original global optimum is mapped to the factored space, how other types of critical points are introduced, and what are their properties.  To answer these questions and conclude from the previous two examples,  we require that the function $f(\mX)$ in~\eqref{nvx:eqn:P0}-\eqref{nvx:eqn:P1}  be
restricted well-conditioned\footnote{Note that the constant $1.5$ for the dynamic range $\frac{\beta}{\alpha}$ in~\eqref{nvx:eqn:assumption} is not optimized and it is possible to slightly relax this constraint with more sophisticated analysis. However, the example of the weighted PCA in~\eqref{nvx:eqn:pca:g:2} implies that the room for improving this constant is rather limited. In particular, Claim~\ref{nvx:claim} and~\eqref{nvx:eqn:wpca:q} indicate that when $\frac{\beta}{\alpha} >3 $, the spurious local minima will occur for the weighted PCA in~\eqref{nvx:eqn:pca:g:2}. Thus, as a sufficient condition for any general objective function to have no spurious local minima, a universal bound on the condition number should be at least no larger than 3, i.e., $\frac{\beta}{\alpha}\leq 3$. Also aside from the lack of spurious local minima, as stated in Theorem~\ref{nvx:thm:main}, the strict saddle property is the other one that needs to be guaranteed.}:
\begin{align}
&\alpha\|\mD\|_F^2\leq [\nabla^2 f(\mX)](\mD,\mD)\leq \beta \|\mD\|_F^2 \text{ with }\ \beta/\alpha\leq1.5 \text{ whenever } \tmop{rank}({\mX})\leq 2r\text{ and }\rank(\mD)\leq 4r.  \tag{$\mathcal{C}$} \label{nvx:eqn:assumption}
\end{align}
We show that as long as the function $f(\mX)$ in the original convex programs satisfies the restricted well-conditioned assumption~\eqref{nvx:eqn:assumption}, each critical point of the factored programs either corresponds to the low-rank globally optimal solution of the original convex programs or is a strict saddle point where the Hessian matrix $\nabla^2 g$  has a strictly negative eigenvalue. This nice geometric structure coupled with the powerful algorithmic tools provided in Theorem \ref{nvx:thm:informal} thus allows simple iterative algorithms to solve the factored programs to a global optimum.

\begin{theorem}[\bf Informal statement of our results]\label{nvx:thm:main}
	Suppose the objective function $f(\mX)$  satisfies the restricted well-conditioned assumption~\eqref{nvx:eqn:assumption}.
	Assume $\mX^\star$ is an optimal solution of~\eqref{nvx:eqn:P0} or~\eqref{nvx:eqn:P1} with $\rank(\mX^\star)= r^\star$. Set $r\geq r^\star$ for the factored variables $\mU$ and $\mV$.
	Then any critical point $\mU$ (or $(\mU,\mV)$) of the factored objective function $g$ in~\eqref{nvx:eqn:F0}-\eqref{nvx:eqn:F1} either corresponds to the global optimum $\mX^\star$ such that $\mX^\star=\mU\mU^\top $ for~\eqref{nvx:eqn:P0} (or $\mX^\star=\mU\mV^\top $ for~\eqref{nvx:eqn:P1})
	or is a strict saddle point (which includes a local maximum) of   $g$.
\end{theorem}
First note that our result covers both over-parameterization where $r > r^\star$ and exact parameterization where $r = r^\star$, while most existing results in low-rank matrix optimization problems ~\cite{nvx:ge2016matrix,nvx:ge2017no,nvx:li2016symmetry}  mainly consider the exact-parameterization case, i.e.,  $r = r^\star$, due to the hardness of fulfilling the gap between the metric in the factored space and the one in the lifted space for the over-parameterization case. The geometric property established in the theorem ensures that many iterative algorithms~\cite{nvx:ge2015escaping,nvx:sun2016nonconvex,nvx:lee2016gradient} converge to a square-root factor (or a factorization) of $\mX^\star$, even with random initialization. Therefore, we can recover the rank-$r^\star$ global minimizer $\mX^\star$ of~\eqref{nvx:eqn:P0}-\eqref{nvx:eqn:P1} by running local-search algorithms on the factored function $g(\mU)$ (or $g(\mU,\mV)$) if we know an upper bound on the rank $r^\star$.
For problems with additional linear constraints, such as those studied in~\cite{nvx:burer2003nonlinear}, one can combine the original objective function with a least-squares term that penalizes the deviation from the linear constraints. As long as the penalization parameter is large enough, the solution is equivalent to that of the constrained minimization problems and hence is also covered by our result.

\subsection{Stylized Applications}
Our main result only relies on the restricted well-conditionedness  of $f(\mX)$. Therefore, in addition to low-rank matrix recovery problems~\cite{nvx:ge2016matrix,nvx:ge2017no,nvx:li2016symmetry,nvx:zhu2017global,nvx:tu2015low}, it is also applicable to many other low-rank matrix optimization problems with non-quadratic objective functions, including $1$-bit matrix recovery,  robust PCA~\cite{nvx:ge2017no}, and low-rank matrix recovery with non-Gaussian  noise~\cite{nvx:wellner2012log}. For ease of exposition, we list the following stylized applications regarding the PSD matrices. But we note that the results listed below also hold for the cases where $\mX$ are general nonsymmetric matrices.

	\subsubsection{Weighted PCA}
	We already know that in the two-dimensional case, the landscape for the factored weighted PCA problem is  closely related with the dynamic range of the weighting matrix. Now we exploit Theorem \ref{nvx:thm:main} to derive the result for the high-dimensional case.   Consider the  \emph{symmetric} weighted PCA problem where the goal is to recover the ground-truth $\mX^\star$ from a pointwisely-weighted observation  $\mY=\mW\odot \mX^\star$. Here $\mW \in\R^{n\times n}$ is the known weighting matrix and the desired solution $\mX^\star\succeq0$ is of rank $r^\star$. A natural approach is to minimize the following squared $\ell_2$ loss:
	\begin{align}\label{nvx:eqn:wpca}
	\minimize_{\mU\in\R^{n\times r}} \frac{1}{2}\|\mW\odot(\mU\mU^\top-\mX^\star)\|_F^2.
	\end{align}
	Unlike the low-rank approximation problem where $\mW$ is the all-ones matrix, in general there is no analytic solutions for the weighted PCA problem~\eqref{nvx:eqn:wpca} ~\cite{nvx:srebro2003weighted} and directly solving this traditional $\ell_2$ loss~\eqref{nvx:eqn:wpca} is known to be NP-hard~\cite{nvx:gillis2011low}.
	We now apply Theorem \ref{nvx:thm:main} to the weighted PCA problem and show the objective function in~\eqref{nvx:eqn:wpca} has  nice geometric structures. Towards that end, define $f(\mX)=\frac{1}{2}\|\mW\odot(\mX-\mX^\star)\|_F^2$ and compute its directional curvature as
	\[[\nabla^2 f(\mX)](\mD,\mD)=\|\mW\odot \mD\|_F^2.\]
	Since $\beta/\alpha$ is a restricted condition number (conditioning on directions of low-rank matrices), which must be no larger than the standard condition number $ {\lambda_{\max}(\nabla^2 f(\mX))}/{\lambda_{\min}(\nabla^2 f(\mX))}$. Thus, together with~\eqref{nvx:eqn:wpca:q}, we have
	\begin{align*}
	\frac{\beta}{\alpha} \leq \frac{\lambda_{\max}(\nabla^2 f(\mX))}{\lambda_{\min}(\nabla^2 f(\mX))}\leq \frac{\max W_{ij}^2}{\min W_{ij}^2}.
	\end{align*}
	Now we apply Theorem \ref{nvx:thm:main} to characterize the geometry of the factored problem of~\eqref{nvx:eqn:wpca}.
	\begin{corollary}
		Suppose the weighting matrix $\mW$ has a small dynamic range $\frac{\max W_{ij}^2}{\min W_{ij}^2}\leq 1.5$.     Then the objective function of~\eqref{nvx:eqn:wpca} with $r\geq r^\star$ satisfies the strict saddle property and has no spurious local minima.
	\end{corollary}

	\subsubsection{Matrix Sensing}
	We now consider the matrix sensing problem which is presented before in Section~\ref{nvx:sec:enlightening examples}. To apply Theorem \ref{nvx:thm:main}, we first compare the RIP~\eqref{nvx:RIP} with our
	restricted well-conditionedness~\eqref{nvx:eqn:assumption}, which is copied below
	\[
	\alpha\|\mD\|_F^2\leq [\nabla^2 f(\mX)](\mD,\mD)\leq \beta \|\mD\|_F^2 \text{ with }\ \beta/\alpha\leq1.5 \text{ whenever } \tmop{rank}({\mX})\leq 2r\text{ and }\rank(\mD)\leq 4r.
	\]
	Clearly, the restricted well-conditionedness~\eqref{nvx:eqn:assumption} would hold if  the linear measurement operator $\A$ satisfies  the $4r$-RIP with a constant $\delta_r$ such that
	\[\frac{1+\delta_{4r}}{1-\delta_{4r}}\leq 1.5 \iff \delta_{4r}\in\left[0,\frac{1}{5}\right].\]
	Now we can apply Theorem \ref{nvx:thm:main} to characterize the geometry of the following matrix sensing problem after the factored parameterization:
	\begin{align}\label{nvx:eqn:application:sensing}
	\minimize_{\mU\in\R^{n\times r}} \frac{1}{2}\|\y-\A(\mU\mU^\top)\|_2^2.
	\end{align}
	
	\begin{corollary}
		Suppose the linear map $\A$ satisfies the $4r$-RIP~\eqref{nvx:RIP} with $\delta_{4r}\in[0,1/5]$.    Then the objective function of~\eqref{nvx:eqn:application:sensing}  with $r\geq r^\star$ satisfies the strict saddle property and has no spurious local minima.
	\end{corollary}

	\subsubsection{1-bit Matrix Completion}
	1-bit matrix completion, as its name indicates, is the inverse problem of completing a low-rank matrix from a set of 1-bit quantized measurements
	\[Y_{ij} =   \tmop{bit}(X^\star_{ij})\quad \text{ for }(i,j)\in\Omega.\]
	Here, $\mX^\star\in\R^{n\times n}$ is the low-rank PSD matrix of rank $r^\star$,  $\Omega$ is a subset of the indices $[n]\times[n]$, and $\tmop{bit}(\cdot)$ is the 1-bit quantifier which  outputs 0 or 1 in a probabilistic manner:
	\[\tmop{bit}(x)=\begin{cases} 1, &\text{with probability }\sigma(x),\\
	0,  &\text{with probability }1-\sigma(x). \end{cases}\]
	One typical choice for $\sigma(x)$ is the sigmoid function $\sigma(x) = \frac{e^x}{1+e^x}$. To recover $\mX^\star$, the authors of~\cite{nvx:davenport20141} propose to minimizing the negative log-likelihood function
	\begin{align}
	\minimize_{\mX\succeq 0} f(\mX) := -\sum_{(i,j)\in\Omega} \bigg[Y_{ij} \log(\sigma(X_{ij})) + (1-Y_{ij}) \log(1-  \sigma(X_{ij}))\bigg]
	\label{nvx:eqn:1bit}
	\end{align}
	and show that if $\|\mX^\star\|_*\leq c n\sqrt{r^\star}$, $\max_{ij}|\mX^\star_{ij}|\leq c$ for some small constant $c$, and $\Omega$ follows certain random binomial model, solving the minimization of the negative log-likelihood function with some nuclear-norm constraint  would be very likely to produce a satisfying approximation to $\mX^\star$~\cite[Theorem 1]{nvx:davenport20141}.
	
	However, when $\mX^\star$ is  extremely high-dimensional (which is the typical case in practice), it is not efficient to deal with the nuclear norm constraint and hence we propose to minimize the factored formulation of~\eqref{nvx:eqn:1bit}
	\begin{align}
	\minimize_{\mU\in\R^{n\times r}} g(\mU) := -\sum_{(i,j)\in\Omega} \bigg[Y_{ij} \log(\sigma((\mU\mU^\top)_{ij})) + (1-Y_{ij}) \log(1-  \sigma((\mU\mU^\top)_{ij}))\bigg].
	\label{nvx:eqn:1bit:F}
	\end{align}
	In order to utilize Theorem \ref{nvx:thm:main} to understand the landscape of the factored objective function~\eqref{nvx:eqn:1bit:F}, we then check the following directional Hessian quadratic from of $f(\mX)$
	\[[\nabla^2 f(\mX)](\mD,\mD) = \sum_{(ij)\in\Omega} \sigma'(X_{ij}) D_{ij}^2.\]
	For simplicity, consider the case where $\Omega=[n]\times[n]$, i.e., observe full  quantized measurements. This will not increase the acquisition cost too much, since each measurement is of 1 bit. Under this assumption, we  have
	\[\min \sigma'(X_{ij})\|\mD\|_F^2\leq [\nabla^2 f(\mX)](\mD,\mD)\leq \max \sigma'(X_{ij})\|\mD\|_F^2 \quad\Leftrightarrow\quad \frac{\beta}{\alpha} \leq  \frac{\max \sigma'(X_{ij})}{\min \sigma'(X_{ij})}\]
	\begin{lemma}
		Let $\Omega=[n]\times[n].$
		Assume $\|\mX\|_\infty:=\max|X_{i,j}|$ is bounded by $1.3169$. Then the negative log-likelihood function~\eqref{nvx:eqn:1bit} $f(\mX)$ satisfies the restricted well-conditioned property.
	\end{lemma}
	\begin{proof}
		First of all, we claim $\sigma(x)$ is an even, positive function and decreasing when $x\geq0$. This is because the sigmoid function $\sigma(x)$ is odd, $\sigma'(x)=\sigma(x)(1-\sigma(x))>0$ by $\sigma(x)\in(0,1)$, and  $\sigma''(x)=-\frac{e^x \left(e^x-1\right)}{\left(e^x+1\right)^3}<0$ for $x\geq0$.  Therefore, for any $|X_{ij}|\leq 1.3169,$  we have $\frac{\max \sigma'(X_{ij})}{\min \sigma'(X_{ij})}= \frac{\max \sigma'(0)}{\min \sigma'(1.3169)}\leq 1.49995\leq 1.5.$
	\end{proof}
	We now use Theorem \ref{nvx:thm:main} to characterize the landscape  of the factored formulation~\eqref{nvx:eqn:1bit:F} in the set $\mathcal{B}_\mU:=\{\mU\in\R^{n\times r}:\|\mU\mU^\top\|_\infty\leq 1.3169.\}$
	
	\begin{corollary}
		Set $r\geq r^\star$ in ~\eqref{nvx:eqn:1bit:F}.  Then the objective function~\eqref{nvx:eqn:1bit:F} satisfies the strict saddle property and has no spurious local minima in $\mathcal{B}_\mU.$
	\end{corollary}
	We remark that such a constraint on $\|\mX\|_\infty$ is also required in the seminal work~\cite{nvx:davenport20141}, while by using the Burer-Monteiro parameterization, our result removes the time-consuming nuclear norm constraint.

	\subsubsection{Robust PCA}
	For the symmetric variant of robust PCA, the observed matrix $\mY=\mX^\star+\mS$ with $\mS$ being sparse and $\mX^\star$ being PSD. Traditionally, we recover $\mX^\star$ by minimizing $||\mY-\mX||_1=\sum_{ij} |Y_{ij}-X_{ij}|$ subject to a PSD constraint. However, this formulation does not directly fit into our framework due to the non-smoothness of the $\ell_1$ norm. An alternative approach is to minimize $\sum_{ij} h_a(Y_{ij}-X_{ij})$, where $h_a(.)$ is chosen to be a convex smooth approximation to the absolute value function. A possible choice is $h_a(x)=a \log((\exp(x/a)+\exp(-x/a))/2)$, which is shown to be strictly convex and smooth in~\cite[Lemma A.1]{nvx:sun2015complete}.

	\subsubsection{Low-rank Matrix Recovery with Non-Gaussian Noise}
	Consider the PCA problem where the underlying noise is non-Gaussian:
	\begin{align*}
	\mY=\mX^\star+\mZ,
	\end{align*}
	i.e., the noise matrix $\mZ\in\R^{n\times n}$ may not follow the Gaussian distributions. Here, $\mX^\star\in\R^{n\times n}$ is a PSD matrix of rank $r^\star$.
	It is known that when the noise is  from normal distribution, the according maximum likelihood estimator (MLE) is given by the minimizer of a squared loss function $\minimize_{\mX\succeq0} \frac{1}{2}\|\mY-\mX\|_F^2.$
	However, in practice, the  noise is often from other distributions~\cite{nvx:sciacchitano2017image}, such as  Poisson,  Bernoulli, Laplacian, and  Cauchy, just to name a few.   In these cases, the resulting MLE, obtained by minimizing the negative log-likelihood function,  is not the square loss one. Such a noise-adaptive estimator is more effective than square-loss minimization. To have a strongly convex and smooth objective function, the noise distribution should be log-strongly-concave, e.g., the Subbotin densities~\cite[Example 2.13]{nvx:wellner2012log},  the Weibull density $f_{\beta}(x)=\beta x^{\beta-1}\tmop{exp}(-x^{\beta})$ for $\beta\geq 2$ ~\cite[Example 2.14]{nvx:wellner2012log}, and the Chernoff's density ~\cite[Conjecture 3.1]{nvx:balabdaoui2014chernoff}. Once the restricted well-conditioned assumption~\eqref{nvx:eqn:assumption} is satisfied, we can then apply Theorem \ref{nvx:thm:main} to characterize the landscape of the factored formulation. Similar results apply to matrix sensing and weighted PCA when the underlying noise is non-Gaussian.

\subsection{Prior Arts and Inspirations}

\paragraph{Prior Arts in Non-convex Optimization Problems}
The past few years have seen a surge of interest in non-convex reformulations of convex optimization problems for efficiency and scalability reasons. However, fully understanding this phenomenon, mainly the landscapes of these non-convex reformulations could be hard. Even certifying the local optimality of a point might be an NP-hard problem~\cite{nvx:murty1987some}. The existence of spurious local minima that are not global optima is a common issue~\cite{nvx:sontag1989backpropagation,nvx:eddy1998profile}. Also, degenerate saddle points or those surrounded by plateaus of small curvature could also prevent local-search algorithms from converging quickly to local optima~\cite{nvx:dauphin2014identifying}.
Fortunately, for a range of convex optimization problems, particularly those involving low-rank matrices, the corresponding non-convex reformulations have nice geometric structures that allow local-search algorithms to converge to global optimality. Examples include low-rank matrix factorization, completion and sensing  ~\cite{nvx:ge2016matrix,nvx:ge2017no,nvx:li2016symmetry,nvx:zhu2017global}, tensor decomposition and completion~\cite{nvx:ge2015escaping,nvx:anandkumar2014guaranteed},
dictionary learning~\cite{nvx:sun2015complete},
phase retrieval~\cite{nvx:sun2016geometric}, and many more.
Based on whether smart initializations are needed, these previous works can be roughly classified into two categories. In one case, the algorithms require a problem-dependent initialization plus local refinement.
A good initialization can lead to global convergence if the initial iterate lies in the attraction basin of the global optima~\cite{nvx:bhojanapalli2015dropping,nvx:sun2015guaranteed,nvx:anandkumar2014guaranteed,nvx:candes2015phase}. For low-rank matrix recovery problems, such initializations can be obtained using spectral methods~\cite{nvx:bhojanapalli2015dropping,nvx:sun2015guaranteed}; for other problems, it is more difficult to find an initial point located in the attraction basin~\cite{nvx:anandkumar2014guaranteed}. The second category of works attempt to understand the empirical success of simple algorithms such as gradient descent~\cite{nvx:lee2016gradient}, which converge to global optimality even with random initialization~\cite{nvx:lee2016gradient,nvx:ge2015escaping,nvx:ge2016matrix,nvx:ge2017no,nvx:li2016symmetry,nvx:zhu2017global}. This is achieved by analyzing the objective function's landscape and showing that they have no spurious local minima and no degenerate saddle points. Most of the works in the second category are for specific matrix sensing problems with quadratic objective functions. Our work expands this line of geometry-based convergence analysis by considering low-rank matrix optimization problems with general objective functions.

\paragraph{\bf Burer-Monteiro Reformulation for PSD Matrices}
In~\cite{nvx:bhojanapalli2015dropping}, the authors also considered low-rank and PSD matrix optimization problems with general objective functions. They characterized the local landscape around the global optima, and hence their algorithms require proper initializations for global convergence. We instead characterize the global landscape by categorizing all critical points into global optima and strict saddles. This guarantees that several local-search algorithms with random initialization will converge to the global optima. Another closely related work is low-rank and PSD matrix recovery from linear observations by minimizing the factored quadratic objective function~\cite{nvx:bhojanapalli2016lowrankrecoveryl}. Low-rank matrix recovery from linear measurements is a particular case of our general objective function framework. Furthermore, by relating the first order optimality condition of the factored problem with the global optimality of the original convex program, our work provides a more transparent relationship between geometries of these two problems and dramatically simplifies the theoretical argument.
More recently, the authors of~\cite{nvx:boumal2016non} showed that for general SDPs with linear objective functions and linear constraints, the factored problems have no spurious local minimizers. In addition to showing non-existence of spurious local minimizers for general objective functions, we also quantify the curvature around the saddle points, and our result covers both over and exact parameterizations.

\paragraph{\bf Burer-Monteiro Reformulation for General Matrices}
The most related work is nonsymmetric matrix sensing from linear observations, which minimizes the factored quadratic objective function~\cite{nvx:park2017non}.
The ambiguity in the factored parameterization
$$\mU\mV^\top = (\mU\mR)\left(\mV{\mR^{-1}}^\top\right)^\top  \text{ for all nonsingular }\mR$$
tends to make the factored quadratic objective function badly-conditioned, especially when the matrix $R$ or its inverse is close to being singular. To overcome this problem, the regularizer
\begin{align}\label{nvx:eqn:equal_energy}
\Theta_E(\mU,\mV)=\|\mU^\top \mU-\mV^\top \mV\|_F^2
\end{align}
is proposed to ensure that $\mU$ and $\mV$ have almost equal energy~\cite{nvx:tu2015low,nvx:park2017non,nvx:zhu2017globalOptimality}. In particular, with the regularizer in~\eqref{nvx:eqn:equal_energy}, it was shown in~\cite{nvx:park2017non,nvx:zhu2017globalOptimality} that $\widetilde g(\mU,\mV) = f(\mU\mV^\top) + \mu \Theta_E(\mU,\mV)$ with a properly chosen $\mu>0$ has similar geometric result as the one provided in Theorem~\ref{nvx:thm:informal} for~\eqref{nvx:eqn:P1}, i.e., $\widetilde g(\mU,\mV)$ also obeys the strict saddle property. Compared with~\cite{nvx:tu2015low,nvx:park2017non,nvx:zhu2017globalOptimality}, our result shows that it is not necessary to introduce the extra regularization~\eqref{nvx:eqn:equal_energy} if we solve~\eqref{nvx:eqn:P1} with the factorization approach. Indeed, the optimization form $\|\mX\|_*=\min_{\mX=\mU\mV^\top }(\|\mU\|_F^2+\|\mV\|_F^2)/2$ of the nuclear norm implicitly requires $\mU$ and $\mV$ to have equal energy. On the other hand, we stress that our interest is to analyze the non-convex geometry of the convex problem~\eqref{nvx:eqn:P1} which as we explained before, has a very nice statistical performance such as it achieves minimax denoising rate~\cite{nvx:candes2010matrix}. Our geometrical result implies that instead of using convex solvers to solve~\eqref{nvx:eqn:P1}, one can turn to apply local-search algorithms to solve its factored problem~\eqref{nvx:eqn:F1} efficiently. In this sense, as a reformulation of the convex program~\eqref{nvx:eqn:P1}, the non-convex optimization problem ~\eqref{nvx:eqn:F1} inherits all the statistical performance bounds for~\eqref{nvx:eqn:P1}.
Cabral et al.~\cite{nvx:cabral2013unifying} worked on a similar problem and showed all global optima of~\eqref{nvx:eqn:F1} corresponds to the solution of the convex program~\eqref{nvx:eqn:P1}. The work~\cite{nvx:haeffele2015global} applied the factorization approach to a more broad class of problems. When specialized to matrix inverse problems, their results show that any local minimizer $\mU$ and $\mV$ with zero columns is a global minimum for the over-parameterization case, i.e., $r>\rank(\mX^\star)$. However, there are no results discussing the existence of spurious local minima or the degenerate saddles in these previous works.  We extend these works and further prove that as long as the loss function $f(\mX)$ is restricted well-conditioned, all local minima are global minima and there are no degenerate saddles with no requirement on the dimension of the variables. {We finally note that compared with~\cite{nvx:haeffele2015global}, our result (Theorem~\ref{nvx:thm:main}) does not depend on the existence of zero columns at the critical points and hence can provide guarantees for many local-search algorithms.
}

\subsection{Notations}

Denote $[n]$ as the collection of all positive integers up to $n$. The symbols $\eye$ and  $\zero$ are reserved for the identity matrix and zero matrix/vector, respectively. A subscript is used to indicate its dimension when this is not clear from context.
We call a matrix PSD, denoted by $\mX\succeq0$, if it is symmetric and all its eigenvalues are nonnegative. The notation $\mX\succeq \mY$ means $\mX-\mY\succeq 0$, i.e., $\mX-\mY$ is PSD. The set of $r\times r$ orthogonal matrices is denoted by $\O_r = \{\mR \in \R^{r\times r}: \mR\mR^\top  = \eye_r\}$. Matrix norms, such as the spectral, nuclear, and Frobenius norms, are denoted respectively by $\|\cdot\|$, $\|\cdot\|_*$ and $\|\cdot\|_F$.

The gradient of a scalar function $f(\mZ)$ with a matrix variable $\mZ\in\R^{m\times n}$ is an $m\times n$ matrix, whose $(i,j)$th entry is $[\nabla f(\mZ)]_{i,j}= \frac{\partial f(\mZ)}{\partial Z_{ij}}$ for $i\in [m]$, $j\in[n]$.  Alternatively, we can view the gradient as a linear form $[\nabla f(\mZ)](\mG) = \lg \nabla f(\mZ), \mG\rg = \sum_{i,j}\frac{\partial f(\mZ)}{\partial Z_{ij}} G_{ij}$ for any $\mG \in \R^{m\times n}$. The Hessian of $f(\mZ)$ can be viewed as a $4$th order tensor of dimension $m\times n\times m\times n$, whose $(i,j,k,l)$th entry is $[\nabla^2 f(\mZ)]_{i,j,k,l}=\frac{\partial^2 f(\mZ)}{\partial Z_{ij}\partial Z_{k,l} }$ for $i, k \in [m]$, $j,l\in[n]$. Similar to the linear form representation of the gradient, we can view the Hessian as a bilinear form defined via $[\nabla^2 f(\mZ)](\mG,\mH)=\sum_{i,j,k,l}\frac{\partial^2 f(\mZ)}{\partial Z_{ij}\partial Z_{kl} } G_{ij}H_{kl}$ for any $\mG,\mH\in\R^{m\times n}$. Yet another way to represent the Hessian is as an $mn\times mn$ matrix $[\nabla^2 f(\mZ)]_{i,j}=\frac{\partial^2 f(\mZ)}{\partial z_i\partial z_j}$ for $i,j\in[mn]$, where $z_i$ is the $i$th entry of the vectorization of $\mZ$. We will use these representations interchangeably whenever the specific form can be inferred from context. For example, in the restricted well-conditionedness assumption~\eqref{nvx:eqn:assumption}, the Hessian is apparently viewed as an $n^2\times n^2$ matrix and the identity $\eye$ is of dimension $n^2\times n^2.$

For a matrix-valued function $\phi: \R^{p\times q} \rightarrow \R^{m\times n}$, it is notationally easier to represent its gradient (or Jacobian) and Hessian as multi-linear operators. For example, the gradient, as a linear operator from $\R^{p\times q}$ to $\R^{m\times n}$, is defined via $[\nabla [\phi(\mU)](\mG)]_{ij} =  \sum_{k \in [p],l\in [q]} \frac{\partial [\phi(\mU)]_{ij}}{\partial U_{kl}} G_{kl}$ for $i \in [m], j \in [n]$ and $\mG \in \R^{p\times q}$; the Hessian, as a bilinear operator from $\R^{p\times q}\times \R^{p\times q}$ to $\R^{m\times n}$, is defined via $[\nabla^2 [\phi(\mU)](\mG, \mH)]_{ij} =  \sum_{k_1, k_2 \in [p],l_1, l_2\in [q]} \frac{\partial^2 [\phi(\mU)]_{ij}}{\partial U_{k_1l_1} \partial U_{k_2 l_2}} G_{k_1l_1}H_{k_2l_2}$ for $i \in [m], j \in [n]$ and $\mG, \mH \in \R^{p\times q}$. Using this notation, the Hessian of the scalar function $f(\mZ)$ of the previous paragraph, which is also the gradient of $\nabla f(\mZ) : \R^{m\times n} \rightarrow \R^{m\times n}$, can be viewed as a linear operator from $\R^{m\times m}$ to $\R^{m\times n}$ denoted by $[\nabla^2 f(\mZ)](\mG)$ and satisfies  $\lg [\nabla^2 f(\mZ)](\mG)], \mH\rg = [\nabla^2 f(\mZ)](\mG, \mH)$ for $\mG, \mH \in \R^{m\times n}$.

\section{Problem Formulation}
This work considers two problems: (i) the minimization of a general convex function $f(\mX)$ with the domain being positive semi-definite matrices; (ii) the minimization of a general convex function $f(\mX)$ regularized by the matrix nuclear norm $\|\mX\|_*$ with the domain being general matrices. Let $\mX^\star$ be an optimal solution of~\eqref{nvx:eqn:P0} or~\eqref{nvx:eqn:P1} of rank $r^\star$. To develop faster and scalable algorithms,
we apply Burer-Monteiro style parameterization~\cite{nvx:burer2003nonlinear} to the low-rank optimization variable $\mX$ in~\eqref{nvx:eqn:P0}-\eqref{nvx:eqn:P1}:
\begin{align*}
\text{ For  symmetric case: }\quad&X =\phi(\mU) := \mU\mU^\top  \\
\text{ For nonsymmetric case: } \quad& X = \psi(\mU,\mV) := \mU\mV^\top
\end{align*}
where $\mU \in \R^{n\times r}$ and $\mV \in\R^{m\times r}$ with $r\geq r^\star$.
With the optimization variable $\mX$ being parameterized, the convex programs are transformed into the factored problems ~\eqref{nvx:eqn:F0}-\eqref{nvx:eqn:F1}:
\begin{align*}
\text{ For  symmetric case: } \quad&\minimize_{\mU \in \R^{n\times r}} g(\mU)=f(\phi(\mU))  \\
\text{ For nonsymmetric case: } \quad& \minimize_{\mU \in \R^{n\times r},V\in\R^{m\times r}} g(\mU,\mV)=f(\psi(\mU,\mV))+ \frac{\lambda}{2}\left(\|\mU\|_F^2+\|\mV\|_F^2\right)
\end{align*}
Inspired by the lifting technique in constructing SDP relaxations, we refer to the variable $\mX$ as the lifted variable, and the variables $\mU,\mV$ as the factored variables. Similar naming conventions apply to the optimization problems, their domains, and objective functions.

\subsection{Consequences of the Restricted Well-conditionedness Assumption}

First the  restricted well-conditionedness assumption reduces to~\eqref{nvx:RIP} when the objective function is quadratic. Moreover, the restricted well-conditioned assumption~\eqref{nvx:eqn:assumption}
shares a similar spirit with~\eqref{nvx:RIP} in
that the operator $\frac{2}{\beta+\alpha} [\nabla^2 f(\mX) ]$ preserves geometric structure for low-rank matrices:
\begin{proposition}\label{nvx:pro:RIP}
	Let $f(\mX)$ satisfy the restricted well-conditionedness assumption~\eqref{nvx:eqn:assumption}. Then
	\begin{align}\label{nvx:eqn:pro:rip}
	&\left|\frac{2}{\beta+\alpha}[\nabla^2f(\mX)](\mG,\mH) - \langle \mG,\mH \rangle\right| \leq \frac{\beta-\alpha}{\beta+\alpha}\|\mG\|_F \|\mH\|_F\leq\frac{1}{5}\|\mG\|_F \|\mH\|_F
	\end{align}
	for any  matrices $\mX,\mG,\mH$ of rank at most $2r$.
\end{proposition}
\begin{proof}
	We extend the argument in~\cite{nvx:candes2008restricted} to a general function $f(\mX)$.
	If either $\mG$ or $\mH$ is zero,~\eqref{nvx:eqn:pro:rip} holds since both sides are $0$.
	For nonzero $\mG$ and $\mH$, we can assume $\|\mG\|_F = \|\mH\|_F = 1$ without loss of generality\footnote{Otherwise, we can divide both sides of the equation~\eqref{nvx:eqn:pro:rip} by $\|\mG\|_F \|\mH\|_F $ and use the homogeneity to get an equivalent version of Proposition \ref{nvx:pro:RIP} with $\mG= \mG/\|\mG\|_F$ and $\mH= \mH/\|\mH\|_F$, i.e., $\|\mG\|_F=\|\mH\|_F=1$.}. Then the assumption~\eqref{nvx:eqn:assumption} implies
	\begin{align*}
	&\alpha \left\|\mG-\mH\right\|_F^2 \leq [\nabla^2 f(\mX)](\mG-\mH,\mG-\mH) \leq \beta \left\|\mG-\mH\right\|_F^2, \\
	&\alpha \left\|\mG+\mH\right\|_F^2 \leq [\nabla^2 f(\mX)](\mG+\mH,\mG+\mH) \leq \beta \left\|\mG+\mH\right\|_F^2.
	\end{align*}
	Thus we have
	\begin{align*}
	\left|2\left[\nabla^2f(\mX)\right](\mG,\mH) - (\beta+\alpha)\left\langle \mG,\mH \right\rangle
	\right|
	&\leq \frac{\beta-\alpha}{2} \underbrace{\left(\left\|\mG\right\|_F^2 +\left\|\mH\right\|_F^2\right)}_{=2}
	= \beta-\alpha=(\beta-\alpha)\underbrace{\|\mG\|_F\|\mH\|_F}_{=1}.
	\end{align*}
	We complete the proof by dividing both sides by $\beta+\alpha$:
	\[\left|\frac{2}{\beta+\alpha}[\nabla^2f(\mX)](\mG,\mH) - \langle \mG,\mH \rangle\right| \leq \frac{\beta-\alpha}{\beta+\alpha}\|\mG\|_F \|\mH\|_F\leq \frac{\beta/\alpha-1}{\beta/\alpha+1} \|\mG\|_F \|\mH\|_F\leq\frac{1}{5}\|\mG\|_F \|\mH\|_F,\]
	where in the last inequality we use the assumption that $\beta/\alpha\leq 1.5.$
\end{proof}

Another immediate consequence of this assumption is that if the original convex program~\eqref{nvx:eqn:P0} has an optimal solution $\mX^\star$  with $\rank(\mX^\star)\leq r$, then there is no other optimum of~\eqref{nvx:eqn:P0} of rank less than or equal to $r$:
\begin{proposition}\label{nvx:pro:1}
	Suppose the function $f(\mX)$  satisfies  the restricted well-conditionedness~\eqref{nvx:eqn:assumption}.
	Let $\mX^\star$ be an optimum of~\eqref{nvx:eqn:P0} with $\rank(\mX^\star)\leq r$. Then $\mX^\star$ is the unique global optimum of~\eqref{nvx:eqn:P0} of rank at most $r$.
\end{proposition}

\begin{proof}
	For the sake of a contradiction, suppose there exists another optimum $\mX$ of~\eqref{nvx:eqn:P0} with $\rank(\mX)\leq r$ and $\mX\neq \mX^\star$. We begin with the second order Taylor expansion, which reads
	\begin{align*}
	f(\mX)=f(\mX^\star)+\lg \nabla f(\mX^\star), \mX-\mX^\star\rg+ \frac{1}{2}[\nabla^2 f(t \mX^\star+ (1-t)\mX)](\mX-\mX^\star,\mX-\mX^\star),
	\end{align*}
	for some $t\in[0,1]$.
	The KKT conditions for the convex optimization problem~\eqref{nvx:eqn:P0}  states that $\nabla f(\mX^\star)\succeq 0$ and $\nabla f(\mX^\star) \mX^\star=\zero$, implying that the second term in the above Taylor expansion
	\[\lg \nabla f(\mX^\star), \mX-\mX^\star\rg=\lg \nabla f(\mX^\star), \mX\rg\geq 0,\]
	since $\mX$ is feasible and hence PSD.
	Further, since $\rank(t \mX^\star+ (1-t)\mX) \leq\rank(\mX)+\rank(\mX^\star)\leq 2r$ and similarly $\rank(\mX-\mX^\star)\leq 2r < 4r$, then from the restricted well-conditionedness assumption~\eqref{nvx:eqn:assumption}  we have
	$$[\nabla^2 f(\tilde \mX)](\mX-\mX^\star,\mX-\mX^\star)\geq \alpha\|\mX-\mX^\star\|_F^2.$$
	Combining all, we obtain a contradiction when $\mX\neq \mX^\star$:
	\begin{align*}
	f(\mX)&\geq f(\mX^\star)+ \frac{1}{2}\alpha\|\mX-\mX^\star\|_F^2
	\geq f(\mX)+ \frac{1}{2}\alpha\|\mX-\mX^\star\|_F^2
	>f(\mX).
	\end{align*}
	where the second inequality follows from the optimality of $\mX^\star$ and the third inequality holds for any $\mX\neq \mX^\star$.
\end{proof}
At a high-level, the proof essentially depends on the restricted strongly convexity of the objective function of the convex program~\eqref{nvx:eqn:P0}, which is guaranteed by the restricted well-conditionedness assumption~\eqref{nvx:eqn:assumption} on $f(\mX)$. The similar argument holds for  ~\eqref{nvx:eqn:P1} by noting that
the sum of a  (restricted) strongly convex function and a standard convex function is still (restricted) strongly convex.
However,   showing this requires a slightly more complicated argument due to the non-smoothness of $\|\mX\|_*$ around those nonsingular matrices. Mainly, we need to use the concept of subgradient.

\begin{proposition}\label{nvx:pro:1:b}
	Suppose the function $f(\mX)$  satisfies the restricted well-conditionedness  ~\eqref{nvx:eqn:assumption}.
	Let $\mX^\star$ be a global optimum of
	\eqref{nvx:eqn:P1} with $\rank(\mX^\star)\leq r$. Then $\mX^\star$ is the unique global optimum of~\eqref{nvx:eqn:P1} of rank at most $r$.
\end{proposition}

\begin{proof}
	For the sake of contradiction, suppose that there exists another optimum $\mX$ of~\eqref{nvx:eqn:P1} with $\rank(\mX)\leq r$ and $\mX\neq \mX^\star$. We begin with the second order Taylor expansion of $f(\mX)$, which reads
	\begin{align*}
	f(\mX)&=f(\mX^\star)+\lg \nabla f(\mX^\star), \mX-\mX^\star\rg+ \frac{1}{2}[\nabla^2 f(t \mX^\star+ (1-t)\mX)](\mX-\mX^\star,\mX-\mX^\star)
	\end{align*}
	for some $t\in[0,1]$.
	From the convexity of $\|\mX\|_*$, for any $\mD\in\partial \|\mX^\star\|_*$, we also have
	\begin{align*}
	\|\mX\|_*&\geq \|\mX^\star\|_*+\lg \mD, \mX-\mX^\star\rg.
	\end{align*}
	Combining both, we obtain
	\begin{align*}
	f(\mX)+\lambda\|\mX\|_*
	&\stack{\ding{172}}{\geq} f(\mX^\star)+\lambda\|\mX^\star\|_*+\lg \nabla f(\mX^\star)+\lambda \mD, \mX-\mX^\star\rg+\frac{1}{2}[\nabla^2 f(t \mX^\star+ (1-t)\mX)](\mX-\mX^\star,\mX-\mX^\star)\\
	&\stack{\ding{173}}{\geq}  f(\mX^\star)+\lambda\|\mX^\star\|_*+\frac{1}{2}[\nabla^2 f(t \mX^\star+ (1-t)\mX)](\mX-\mX^\star,\mX-\mX^\star)\\
	&\stack{\ding{174}}{\geq}  f(\mX^\star)+\lambda\|\mX^\star\|_*+\frac{1}{2}\alpha\|\mX-\mX^\star\|_F^2\\
	&\stack{\ding{175}}{=}  f(\mX)+\lambda\|\mX\|_*+\frac{1}{2}\alpha\|\mX-\mX^\star\|_F^2\\
	&\stack{\ding{176}}{>}  f(\mX)+\lambda\|\mX\|_*,
	\end{align*}
	where \ding{172} holds for any $\mD\in\partial \|\mX^\star\|_*$.
	For \ding{173}, we use fact that
	$\partial f_1 +\partial f_2 = \partial (f_1+f_2)$ for any convex functions  $f_1,f_2,$ to obtain that $ \nabla f(\mX^\star)+\lambda  \partial \|\mX^\star\|_* =\partial (f(\mX^\star)+\lambda\|\mX^\star\|_*)$, which includes   $\zero$  since $\mX^\star$ is a global optimum of~\eqref{nvx:eqn:P1}. Therefore,  \ding{173} follows by  choosing $\mD\in\partial \|\mX^\star\|_*$ such that $\nabla f(\mX^\star)+\lambda \mD=\zero$.
	\ding{174} uses the restricted well-conditionedness assumption~\eqref{nvx:eqn:assumption} as  $\rank(t \mX^\star+ (1-t)\mX) \leq 2r$ and $\rank(\mX-\mX^\star)\leq 4r$.
	\ding{175} comes from the assumption that both $\mX$ and $\mX^\star$ are global optimal solutions of~\eqref{nvx:eqn:P1}.
	\ding{176}  uses the assumption that $\mX\neq \mX^\star.$
\end{proof}

\section{Understanding the Factored Landscapes for PSD Matrices}
In the convex program~\eqref{nvx:eqn:P0},  we minimize  a convex function $f(\mX)$ over the PSD cone. Let $\mX^\star$ be an optimal solution of~\eqref{nvx:eqn:P0} of rank $r^\star$. We re-parameterize  the low-rank PSD variable $\mX$ as
\[
\mX = \phi(\mU)=\mU\mU^\top \]
where $\mU \in \R^{n\times r}$ with $r \geq r^\star$ is a rectangular, matrix square root of $\mX$. After this parametrization, the convex program is transformed into the factored problem~\eqref{nvx:eqn:F0} whose objective function is $g(\mU) =f(\phi(\mU))$.

\subsection{Transforming the Landscape for PSD Matrices}
Our primary interest is to understand how the landscape of the lifted objective function $f(\mX)$ is transformed by the factored parameterization $\phi(\mU) = \mU\mU^\top $, particularly how its global optimum is mapped to the factored space, how other types of critical points are introduced, and what  their properties are.

We show that if the function $f(\mX)$   is restricted well-conditioned,
then each critical point of the factored objective function $g(\mU)$ in~\eqref{nvx:eqn:F0} either corresponds to the low-rank global solution of the original convex program~\eqref{nvx:eqn:P0} or is a strict saddle where the Hessian $\nabla^2 g(\mU)$ has a strictly negative eigenvalue. This implies that the factored objective function $g(\mU)$ satisfies the strict saddle property.

\begin{theorem}[\bf Transforming the Landscape for PSD matrices]\label{nvx:thm:main:1}
	Suppose the function $f(\mX)$ in~\eqref{nvx:eqn:P0} is  twice continuously differentiable and is  restricted well-conditioned~\eqref{nvx:eqn:assumption}.
	Assume $\mX^\star$ is an optimal solution of ~\eqref{nvx:eqn:P0} with $\rank(\mX^\star)= r^\star$. Set $r\geq r^\star$ in~\eqref{nvx:eqn:F0}.
	Let ${\mU}$ be any critical point of $g(\mU)$ satisfying $\nabla g(\mU)=\zero$. Then $\mU$  either corresponds to a square-root factor of  $\mX^\star$, i.e.,
	\begin{align*}
	\mX^\star=\mU\mU^\top ;
	\end{align*}
	or is a strict saddle of the factored problem ~\eqref{nvx:eqn:F0}.
	More precisely, let $\mU^\star\in\R^{n\times r}$ such that $\mX^\star=\mU^\star \mU^{\star \top}$ and set $\mD=\mU-\mU^\star \mR$ with $\mR=\argmin_{\mR: \mR\in\O_r}\|\mU-\mU^\star \mR\|_F^2$, then the curvature of $\nabla^2 g(\mU)$ along $\mD$ is strictly negative:
	\begin{align*}
	[\nabla^2g(\mU)](\mD,\mD)&\leq
	\begin{cases}
	-0.24\alpha\min\left\{\rho(\mU)^2,\rho(\mX^\star)\right\}\|\mD\|_F^2 & \text{ when $r> r^\star$};\\ \\
	-0.19\alpha\rho(\mX^\star)\|\mD\|_F^2 & \text{ when $r= r^\star$};\\ \\
	-0.24\alpha\rho(\mX^\star)\|\mD\|_F^2 & \text{ when $\mU= \zero$}
	\end{cases}
	\end{align*}
	with $\rho(\cdot)$ denoting the smallest nonzero singular value of its argument. This further implies
	\begin{align*}
	\lambda_{\min}(\nabla^2 g(\mU))&\leq
	\begin{cases}
	-0.24\alpha\min\left\{\rho(\mU)^2,\rho(\mX^\star)\right\} &\text{ when $r> r^\star$};\\ \\
	-0.19\alpha\rho(\mX^\star) &\text{ when $r= r^\star$};\\ \\
	-0.24\alpha\rho(\mX^\star) & \text{ when $\mU= \zero$}.
	\end{cases}
	\end{align*}
\end{theorem}

Several remarks follow. First, the matrix $\mD$ is the direction from the saddle point $\mU$ to its closest globally optimal factor $\mU^\star \mR$ of the same dimension as $\mU$.
Second, our result covers both over-parameterization where $r > r^\star$ and exact parameterization where $r = r^\star$. Third, we can recover the rank-$r^\star$ global minimizer $\mX^\star$ of~\eqref{nvx:eqn:P0} by running local-search algorithms on the factored function $g(\mU)$ if we know an upper bound on the rank $r^\star$. In particular, to apply the results in~\cite{nvx:lee2017first} where the first-order algorithms are proved to escape all the strict saddles, aside from the strict saddle property, one needs  $g(\mU)$ to have a Lipschitz continuous gradient, i.e., $\|\nabla g(\mU) - \nabla g(\mV)\|_F \leq L_c \|U - \mV\|_F$ or $\|\nabla^2 g(\mU)\|\leq L_c$ for some positive constant $L_c$ (also known as the Lipschitz constant). As indicated by the expression of $\nabla^2 g(\mU)$ in~\eqref{nvx:eqn:hess}, it is possible that one can not find such a constant $L_c$ for the whole space. Similar to~\cite{nvx:jin2017escape} which considers the low-rank matrix factorization problem, suppose the local-search algorithm starts at $\mU_0$ and sequentially decreases the objective value (which is true as long as the algorithm obeys certain sufficient decrease property~\cite{nvx:wolfe1969convergence}). Then it is adequate to focus on the sublevel set of $g$
	\begin{align}\label{nvx:eqn:lowset}
	\tmop{Lev}_f({\mU_0})=\left\{U:g(\mU)\leq g(\mU_0)\right\},
	\end{align}
	and show that $g$ has a Lipschitz gradient on $\tmop{Lev}_f({\mU_0})$. This is formally established in Proposition \ref{nvx:pro:4}, whose proof is given in Appendix \ref{nvx:sec:proof:pro:4}.
	\begin{proposition}\label{nvx:pro:4}
		Under the same setting as in Theorem \ref{nvx:thm:main:1}, for any initial point $\mU_0$,  $g(\mU)$ on $\tmop{Lev}_f({\mU_0})$ defined in~\eqref{nvx:eqn:lowset} has a Lipschitz  continuous gradient with the Lipschitz constant
		\[L_c=\sqrt{2\beta \sqrt{\frac{2}{\alpha}(f(\mU_0\mU_0^\top) - f(\mX^\star))} + 2\| \nabla f(\mX^\star) \|_F + 4\beta \left(\|\mU^\star\|_F + \frac{\sqrt{\frac{2}{\alpha}(f(\mU_0\mU_0^\top) - f(\mX^\star))}}{2(\sqrt{2} -1)\rho(\mU^\star)}\right)^2},\]
		where $\rho(\cdot)$ denotes the smallest nonzero singular value of its argument.
	\end{proposition}

\subsection{Metrics in the Lifted and Factored Spaces}
Before continuing this geometry-based argument, it is essential to have a good understanding of the domain of the factored problem and establish a metric for this domain.
Since for any $\mU$, $\phi(\mU) = \phi(\mU\mR)$ where $\mR \in \O_r$, the domain of the factored objective function $g(\mU)$ is stratified into equivalence classes and can be viewed as a quotient manifold~\cite{nvx:absil2009optimization}. The matrices in each of these equivalence classes differ by an orthogonal transformation (not necessarily unique when the rank of $\mU$ is less than $r$). One implication is that, when working in the factored space,
we should consider all factorizations of $\mX^\star:$
\[\A^\star=\{\mU^\star\in\R^{n\times r}: \phi(\mU^\star) = \mX^\star\}.\]
A second implication is that when considering the distance between two points $\mU_1$ and $\mU_2$, one should use the distance between their corresponding equivalence classes:
\begin{align}\label{nvx:eqn:dist:point}
\dist(\mU_1,\mU_2)=\min_{\mR_1\in\O_r,\mR_2\in\O_r}\|\mU_1\mR_1-\mU_2 \mR_2\|_F=\min_{\mR\in\O_r}\|\mU_1-\mU_2 \mR\|_F.
\end{align}
Under this notation, $\dist(\mU,\mU^\star) = \min_{\mR\in\O_r}\|\mU-\mU^\star \mR\|_F$ represents the distance between the class containing a critical point $\mU\in\R^{n\times r}$ and the optimal factor class $\A^\star$.
The second minimization problem in the definition~\eqref{nvx:eqn:dist:point} is known as the orthogonal Procrustes problem, where the global optimum $R$ is characterized by the following lemma:

\begin{lemma}\cite{nvx:higham1995matrix}\label{nvx:lem:procrust}
	An optimal solution for the orthogonal Procrustes problem:
	\[\mR=\argmin_{\tilde{\mR}\in\O_r}\|\mU_1-\mU_2 \tilde{\mR}\|_F^2 = \argmax_{\tilde{\mR}\in\O_r} \lg \mU_1, \mU_2 \tilde{\mR}\rg \]
\end{lemma}

For any two matrices $\mU_1, \mU_2 \in \R^{n\times r}$, the following lemma  relates the distance $\|\mU_1\mU_1^\top -\mU_2\mU_2^\top \|_F$ in the lifted space  to the distance $\dist(\mU_1, \mU_2)$ in the factored space. The proof is deferred to Appendix \ref{nvx:sec:proof:lem:Gongguo}.
\begin{lemma}\label{nvx:lem:Gongguo}
	Assume that $\mU_1, \mU_2\in\R^{n\times r}$.  Then
	\[\|\mU_1\mU_1^\top -\mU_2\mU_2^\top \|_F \geq \min\left\{\rho(\mU_1),\rho(\mU_2)\right\} \dist(\mU_1, \mU_2). \]
\end{lemma}

In particular, when one matrix is of full rank,   we have a similar but tighter result to relate these two distances.
\begin{lemma}~\cite[Lemma 5.4]{nvx:tu2015low} \label{nvx:lem:ben}
	Assume that $\mU_1, \mU_2\in\R^{n\times r}$ and $\rank(\mU_1)=r$.  Then
	\[\|\mU_1\mU_1^\top -\mU_2\mU_2^\top \|_F \geq 2(\sqrt2-1)\rho(\mU_1) \dist(\mU_1, \mU_2). \]
\end{lemma}

\subsection{Proof Idea: Connecting the Optimality Conditions}\label{nvx:sec:QW}
The proof is inspired by connecting the optimality conditions for the two programs~\eqref{nvx:eqn:P0} and~\eqref{nvx:eqn:F0}. First of all, as the critical points of the convex optimization problem~\eqref{nvx:eqn:P0}, they are global optima and are characterized by the necessary and sufficient KKT condition~\cite{nvx:boyd2004convex}
\begin{align}\label{nvx:eqn:optimality}
\nabla f(\mX^\star)\succeq 0, \nabla f(\mX^\star)\mX^\star=\zero, \mX^\star\succeq 0.
\end{align}
The factored optimization problem~\eqref{nvx:eqn:F0} is unconstrained, with the critical points being specified by the zero gradient condition
\begin{align}\label{nvx:eqn:critical}
\nabla g(\mU) = 2\nabla f(\phi(\mU))\mU = \zero.
\end{align}

To classify the critical points of ~\eqref{nvx:eqn:F0}, we compute the Hessian quadratic form $[\nabla^2g(\mU)](\mD,\mD)$ as
\begin{align}\label{nvx:eqn:hess}
[\nabla^2g(\mU)](\mD,\mD)=2\lg\nabla f(\phi(\mU)),\mD\mD^\top \rg +[\nabla^2f(\phi(\mU))](\mD\mU^\top +\mU\mD^\top ,\mD\mU^\top  +\mU\mD^\top ).
\end{align}
Roughly speaking, the Hessian quadratic form has two terms -- the first term involves the gradient of $f(\mX)$ and the Hessian of $\phi(\mU)$, while the second term involves the Hessian of $f(\mX)$ and the gradient of $\phi(\mU)$. Since $\phi(\mU+\mD) = \phi(\mU) + \mU\mD^\top  + \mD\mU^\top  + \mD\mD^\top $, the gradient of $\phi$ is the linear operator $[\nabla \phi(\mU)] (\mD) = \mU\mD^\top  + \mD\mU^\top $ and the Hessian bilinear operator applies as $\frac{1}{2}[\nabla^2 \phi(\mU)](\mD,\mD) = \mD\mD^\top $. Note in~\eqref{nvx:eqn:hess} the second quadratic form is always nonnegative since $\nabla^2 f\succeq 0$ due to the convexity of $f$.

For any critical point $\mU$ of $g(\mU)$, the corresponding lifted variable $\mX:= \mU\mU^\top $ is PSD and satisfies $\nabla f(\mX)\mX = \zero$. On one hand, if $\mX$ further satisfies $\nabla f(\mX) \succeq 0$, then in view of the KKT conditions~\eqref{nvx:eqn:optimality} and noting $\rank(\mX)=\rank(\mU) \leq r$, we must have $\mX = \mX^\star$, the global optimum of~\eqref{nvx:eqn:P0}. On the other hand, if $\mX \neq \mX^\star$, implying $\nabla f(\mX) \nsucceq 0$ due to the necessity of~\eqref{nvx:eqn:optimality}, then additional critical points can be introduced into the factored space. Fortunately, $\nabla f(\mX) \nsucceq 0$ also implies that the first quadratic form in~\eqref{nvx:eqn:hess} might be negative for a properly chosen direction $\mD$. To sum up, the critical points of $g(\mU)$ can be classified into two categories: the global optima in the optimal factor set $\A^\star$ with $\nabla f(\mU\mU^\top ) \succeq 0$ and those with $\nabla f(\mU\mU^\top ) \nsucceq 0$. For the latter case, by choosing a proper direction $\mD$, we will argue that the Hessian quadratic form~\eqref{nvx:eqn:hess} has a strictly negative eigenvalue, and hence moving in the direction of $\mD$ in a short distance will decrease the value of $g(\mU)$, implying that they are strict saddles  and are not local minima.

\begin{figure}[ht!]
	\centering
	\includegraphics[width=0.3\textwidth]{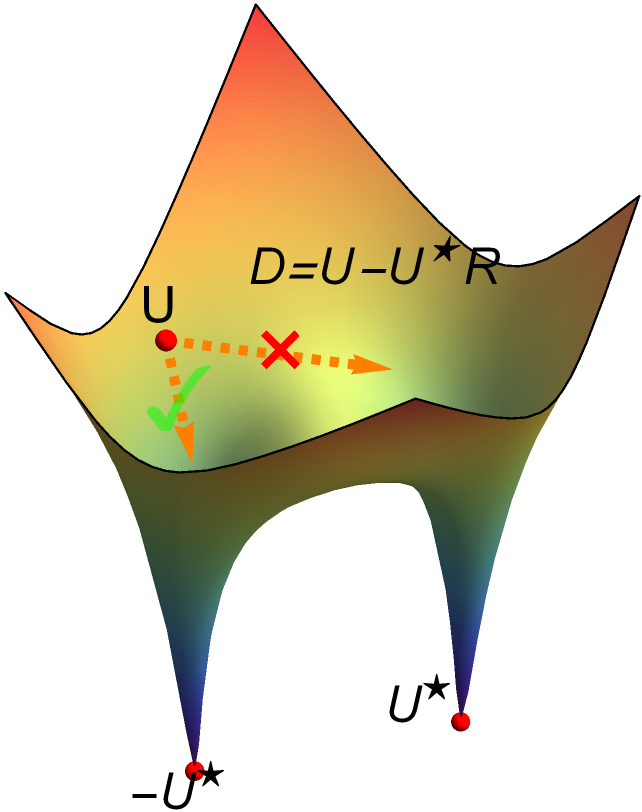}
	\caption{The matrix $\mD=\mU-\mU^\star \mR$ is the direction from the critical point $\mU$ to its nearest optimal factor $\mU^\star \mR$, whose norm $\|\mU-\mU^\star \mR\|_F$ defines the distance  $\dist(\mU,\mU^\star)$. Here, $\mU$ is closer to $-\mU^\star$ than $\mU^\star$ and the direction from $\mU$ to $-\mU^\star$ has more negative curvature compared to the direction from $\mU$ to $\mU^\star$.
	}\label{nvx:fig:distance}
\end{figure}

We argue that a good choice of $\mD$ is the direction from the current $\mU$ to its closest point in the optimal factor set $\A^\star$. Formally, $\mD = \mU-\mU^\star \mR$ where $\mR=\argmin_{\mR:R\in \O_r}\|\mU-\mU^\star \mR\|_F$ is the optimal rotation for the orthogonal Procrustes problem. As illustrated in Figure \ref{nvx:fig:distance} where we have two global solutions $\mU^\star$ and $-\mU^\star$ and $\mU$ is closer to $-\mU^\star$, the direction from $\mU$ to $-\mU^\star$ has more negative curvature compared to the direction from $\mU$ to $\mU^\star$.

Plugging this choice of $\mD$ into the first term of~\eqref{nvx:eqn:hess}, we simplify it as
\begin{align}\label{nvx:eqn:grad:f}
\lg\nabla f(\mU\mU^\top ),\mD\mD^\top  \rg\nn
&=\lg\nabla f(\mU\mU^\top ), \mU^\star \mU^{\star \top}-\mU^\star \mR \mU^\top  - \mU(\mU^\star \mR)^\top +\mU\mU^\top \rg\nonumber \\
& = \lg\nabla f(\mU\mU^\top ), \mU^\star \mU^{\star \top}\rg\nonumber \\
& = \lg\nabla f(\mU\mU^\top ), \mU^\star \mU^{\star \top}-\mU\mU^\top \rg,
\end{align}
where both the second line and last line follow from the critical point property $\nabla f(\mU\mU^\top )\mU = \zero$.
To gain some intuition on why~\eqref{nvx:eqn:grad:f} is negative while the second term in~\eqref{nvx:eqn:hess} remains small, we consider a simple example: the matrix PCA problem.

\paragraph{Matrix PCA Problem.}
Consider the PCA problem for symmetric PSD matrices
\begin{align}
\minimize_{\mX \in \R^{n\times n}} f_{\tmop{PCA}}(\mX):=  \frac{1}{2}\|\mX-\mX^\star\|_F^2\ \st \mX \succeq 0,
\end{align}
where $\mX^\star$ is a symmetric PSD matrix of rank $r^\star$. Trivially, the optimal solution is $\mX = \mX^\star$. Now consider the factored problem
$$\minimize_{\mU\in\R^{n\times r}} g(\mU):= f_{\tmop{PCA}}(\mU\mU^\top )=\frac{1}{2}\|\mU\mU^\top -\mU^\star \mU^{\star \top}\|_F^2,$$
where $\mU^\star\in\R^{n\times r}$ satisfies $\phi(\mU^\star) = \mX^\star$.
Our goal is to show that any critical point $\mU$ such that $\mX:=\mU\mU^\top  \neq \mX^\star$ is a strict saddle.

\paragraph{Controlling the first term.}
Since $\nabla f_{\tmop{PCA}}(\mX)=\mX-\mX^\star$, by~\eqref{nvx:eqn:grad:f}, the first term of $[\nabla^2 g(\mU)](\mD,\mD)$ in~\eqref{nvx:eqn:hess} becomes
\begin{align}\label{nvx:eqn:first:term}
2\lg\nabla f_{\tmop{PCA}}(\mX), \mD\mD^\top \rg
&=2\lg \nabla f_{\tmop{PCA}}(\mX), \mX^\star-\mX\rg
=2\lg \mX-\mX^\star, \mX^\star-\mX\rg
=-2\|\mX-\mX^\star\|_F^2,
\end{align}
which is strictly negative when $\mX\neq \mX^\star$.

\paragraph{Controlling the second term.}
We show that the second term
$[\nabla^2f(\phi(\mU))](\mD\mU^\top +\mU\mD^\top ,\mD\mU^\top  +\mU\mD^\top )$
vanishes by showing that
$\mD\mU^\top =\zero$ (hence $\mU\mD^\top =\zero$). For this purpose, let $\mX^\star = \mQ\diag(\vlambda)\mQ^\top  = \sum_{i=1}^{r^\star} \lambda_i \q_i \q_i^\top $ be the eigenvalue decomposition of $\mX^\star$, where $\mQ = \begin{bmatrix} \q_1 &\cdots & \q_{r^\star} \end{bmatrix} \in \R^{n\times r^\star}$ has orthonormal columns and $\vlambda \in \R^{r^\star}$ is composed of positive entries. Similarly, let $\phi(\mU) = \mV\diag(\boldsymbol \mu)\mV^\top  = \sum_{i=1}^{r'} \mu_i \v_i \v_i^\top $ be the eigenvalue decomposition of $\phi(\mU)$, where $r' = \rank(\mU)$. The critical point $\mU$ satisfies $-\nabla g(\mU)= 2(\mX^\star-\phi(\mU))\mU = \zero$, implying that
\begin{align*}
\zero = \bigg(\mX^\star -\sum_{i=1}^{r'} \mu_i \v_i \v_i^\top \bigg)\v_j = \mX^\star\v_j - \mu_j \v_j, j = 1, \ldots, r'.
\end{align*}
This means $(\mu_j, \v_j)$ forms an eigenvalue-eigenvector pair of $\mX^\star$ for each $j = 1, \ldots, r'$. Consequently,
\[
\mu_j = \lambda_{i_j}\text{ and }\v_j = \q_{i_j},  j = 1, \ldots, r'.
\]
Hence
$\phi(\mU) = \sum_{j=1}^{r'} \lambda_{i_j} \q_{i_j}\q_{i_j}^\top  = \sum_{j=1}^{r^\star} \lambda_j s_j \q_j \q_j^\top $. Here $s_j$ is equal to either 0 or 1 indicating which of the eigenvalue-eigenvector pair $(\lambda_j, \q_j)$ appears in the decomposition of $\phi(\mU)$. Without loss of generality, we can choose $\mU^\star= \mQ\begin{bmatrix} \diag( \sqrt{\vlambda}) & \zero\end{bmatrix}$. Then $\mU=\mQ\begin{bmatrix} \diag( \sqrt{\vlambda}\odot\s) & \zero\end{bmatrix} \mV^\top $ for some orthonormal matrix $\mV \in\R^{r\times r}$ and $\s = \begin{bmatrix}s_1 & \cdots & s_{r^\star}\end{bmatrix}$, where the symbol  $\odot$ means pointwise  multiplication.   By the Procrustes Lemma in\cite{nvx:higham1995matrix}, we obtain $\mR=\mV^\top $. Plugging these into $\mD\mU^\top =\mU\mU^\top -\mU^\star \mR \mU^\top $ gives $\mD\mU^\top  = \zero$.

\paragraph{Combining the two.}
Hence $[\nabla^2g(\mU)](\mD,\mD)$ is simply determined by its first term
\begin{align*}
[\nabla^2g(\mU)](\mD,\mD)
&= -2\|\mU\mU^\top -\mU^\star \mU^{\star \top}\|_F^2\\
&\leq-2\min\left\{\rho(\mU)^2,\rho(\mU^\star)^2\right\}\|\mD\|_F^2 \\
& = -2\min\left\{\rho(\phi(\mU)),\rho(\mX^\star)\right\}\|\mD\|_F^2\\
&= -2\rho(\mX^\star)\|\mD\|_F^2,
\end{align*}
where the second line follows from Lemma \ref{nvx:lem:Gongguo} and  the last line follows from the fact that all the eigenvalues of $\mU\mU^\top $ come from those of $\mX^\star$. Finally, we obtain the desired strict saddle property of $g(\mU)$:
\begin{align*}
\lambda_{\min}(\nabla^2g(\mU))\leq-2\rho(\mX^\star).
\end{align*}

This simple example is ideal in several ways, particularly the gradient $\nabla f(\phi(\mU)) = \phi(\mU) - \phi(\mU^\star)$, which directly establishes the negativity of the first term in~\eqref{nvx:eqn:hess}; and by choosing $\mD=\mU-\mU^\star \mR$ and using $\mD\mU^\top  = \zero$, the second term vanishes. Neither of these simplifications hold for general objective functions $f(\mX)$. However, the example does suggest that the direction $\mD=\mU-\mU^\star \mR$ is a good choice to show $[\nabla^2 g(\mU)](\mD,\mD)\leq -\tau \|\mD\|_F^2 \text{ for some }\tau>0$.
For a formal proof, we will also use the direction $\mD=\mU-\mU^\star \mR$ to show that those critical points $\mU$ not corresponding to $\mX^\star$ have a negative directional curvature for the general factored objective function $g(\mU).$

\subsection{A Formal Proof of Theorem \ref{nvx:thm:main:1}}
\paragraph{Proof Outline.}
We present a formal proof of Theorem \ref{nvx:thm:main:1} in this section. The main argument involves showing each critical point $\mU$ of $g(\mU)$ either corresponds to the optimal solution $\mX^\star$ or its Hessian matrix $\nabla^2 g(\mU)$ has at least one strictly negative eigenvalue. Inspired by the discussions in Section \ref{nvx:sec:QW}, we will use the direction  $\mD=\mU-\mU^\star \mR$ and show that the Hessian $\nabla^2 g(\mU)$ has a strictly negative directional curvature in the direction of $\mD$, i.e., $[\nabla^2 g(\mU)](\mD,\mD)\leq -\tau \|\mD\|_F^2, \text{ for some }\tau>0.$

\paragraph{Supporting Lemmas.}
We first list two  lemmas.  The first lemma separates  $\|(\mU - \mZ ){\mU}^\top  \|_F^2$ into two terms:   $\|\mU\mU^\top   -  \mZ  \mZ^\top   \|_F^2  $ and $\|(\mU\mU^\top   -  \mZ  \mZ^\top ) {\mQ}{\mQ}^\top  \|_F^2$ with $\mQ\mQ^\top $ being the projection matrix onto $\range(\mU)$. It is crucial for  the first term $\|\mU\mU^\top  -  \mZ  \mZ^\top  \|_F^2$  to have a small coefficient. In the second lemma, we will further control the second term as a consequence of $\mU$ being a critical point. The proof of Lemma \ref{nvx:lem:2} is given in Section \ref{nvx:sec:proof:lem2}.

\begin{lemma}
	Let $\mU$ and $ \mZ $ be any two  matrices in $\R^{n\times r}$  such that $\mU^\top    \mZ  =  \mZ^\top   \mU$ is PSD.
	Assume that ${\mQ}$ is an orthogonal matrix whose columns span $\range(\mU)$.
	Then
	\[
	\left\|({\mU} -  \mZ ){\mU}^\top  \right\|_F^2 \leq \frac{1}{8}\left\|\mU\mU^\top   - \mZ  \mZ^\top   \right\|_F^2 + \left(3 + \frac{1}{2\sqrt{2} -2} \right)\left\|(\mU\mU^\top  -  \mZ  \mZ^\top   ) {\mQ}{\mQ}^\top  \right\|_F^2.
	\]
	\label{nvx:lem:2}
\end{lemma}
We remark that Lemma \ref{nvx:lem:2} is a strengthened version of ~\cite[Lemma 4.4] {nvx:bhojanapalli2016lowrankrecoveryl}. While the result there requires: (i) $\mU$ to be a critical point of the factored objective function $g(\mU)$; (ii) $ \mZ $ to be an optimal factor in $\A^\star$ that is closest to $\mU$, i.e., $ \mZ =\mU^\star \mR$ with   $\mU^\star\in\A^\star$ and $\mR=\argmin_{\mR:\mR\mR^\top =\eye_r}\|\mW-\mW^\star \mR\|_F$. Lemma  \ref{nvx:lem:2} removes these assumptions and requires only $\mU^\top    \mZ  =  \mZ^\top   \mU$ being PSD.

Next,  we control the distance between $\mU\mU^\top $  and the global solution $\mX^\star$   when $\mU$ is a critical point of the factored objective function $g(\mU)$, i.e., $\nabla g(\mU)=\zero$. The proof, given in Section \ref{nvx:sec:proof:lem3}, relies on writing $\nabla f(\mX) = \nabla f(\mX^\star)+\int_0^1 [\nabla^2f(t \mX + (1-t)\mX^\star)](\mX-\mX^\star)dt$ and applying
Proposition \ref{nvx:pro:RIP}.
\begin{lemma}[\bf Upper Bound on $\|(\mU\mU^\top -\mU^\star \mU^{\star \top})\mQ\mQ^\top  \|_F^2$]
	Suppose the objective function $f(\mX)$ in~\eqref{nvx:eqn:P0} is  twice continuously differentiable and satisfies the restricted well-conditionedness assumption~\eqref{nvx:eqn:assumption}.
	Further, let $\mU$ be any critical point of~\eqref{nvx:eqn:F0} and  $\mQ$ be the orthonormal basis spanning $\range(\mU)$. Then
	\begin{align*}
	\left\|(\mU\mU^\top -\mU^\star \mU^{\star \top})\mQ\mQ^\top  \right\|_F \leq  \frac{\beta-\alpha}{\beta+\alpha}\left\|\mU\mU^\top -\mU^\star \mU^{\star \top}\right\|_F.
	\end{align*}
	\label{nvx:lem:3}
\end{lemma}

\begin{proof}[\bf Proof of Theorem  \ref{nvx:thm:main:1}]
	Along the same lines as in the matrix PCA example, it suffices to find a direction $\mD$ to produce a strictly negative curvature for each critical point $\mU$ not corresponding to $\mX^\star$. We choose $\mD=\mU-\mU^\star \mR$ where $\mR=\argmin_{\mR:\mR\mR^\top =\eye_r}\|\mW-\mW^\star \mR\|_F$. Then
	\begin{align*}
	&[\nabla^2g(\mU)](\mD,\mD)\\
	=&2\lg\nabla f(\mX),\mD\mD^\top  \rg+ [\nabla^2f(\mX)](\mD\mU^\top +\mU\mD^\top ,\mD\mU^\top  + \mU\mD^\top )\tag*{By Eq.~\eqref{nvx:eqn:hess}}\\
	=&2\lg\nabla f(\mX), \mX^\star-\mX\rg+[\nabla^2f(\mX)](\mD\mU^\top +\mU\mD^\top ,\mD\mU^\top  + \mU\mD^\top )\tag*{By Eq.~\eqref{nvx:eqn:critical}}\\
	\leq&\underbrace{2\lg \nabla f(\mX)-\nabla f(\mX^\star),\mX^\star-\mX\rg}_{\Pi_1}+
	\underbrace{[\nabla^2f(\mX)](\mD\mU^\top +\mU\mD^\top ,\mD\mU^\top  + \mU\mD^\top )}_{\Pi_2}\tag*{By Eq.~\eqref{nvx:eqn:optimality}}
	\end{align*}
	
	In the following, we will bound $\Pi_1$ and $\Pi_2$, respectively.
	\paragraph{Bounding $\Pi_1$.}
	\begin{align*}
	\Pi_1=-2\lg\nabla f(\mX^\star)-\nabla f(\mX), \mX^\star-\mX\rg
	&\stack{\ding{172}}{=}-2\left\lg \int_0^1 [\nabla^2 f(t \mX + (1-t)\mX^\star)](\mX^\star-\mX) d t,\mX^\star-\mX\right\rg\\
	& = -2\int_0^1 \left[\nabla^2 f(t \mX + (1-t)\mX^\star)\right](\mX^\star-\mX, \mX^\star-\mX) d t\\
	& \stack{\ding{173}}{\leq} -2\alpha\|\mX^\star-\mX\|_F^2,
	\end{align*}
	where \ding{172} follows from the Taylor's Theorem for vector-valued functions ~\cite[Eq. (2.5) in Theorem 2.1]{nvx:nocedal2006numerical}, and \ding{173} follows from the restricted strong convexity assumption~\eqref{nvx:eqn:assumption} since the PSD matrix $t \mX+ (1-t)\mX^\star$  has rank of at most $2r$ and $\rank(\mX^\star-\mX)\leq 4r.$

	\paragraph{Bounding $\Pi_2$.}
	\begin{align*}
	\Pi_2&=[\nabla^2f(\mX)](\mD\mU^\top +\mU\mD^\top ,\mD\mU^\top  + \mU\mD^\top )\\
	&\leq \beta\|\mD\mU^\top +\mU\mD^\top \|_F^2\tag*{By ~\eqref{nvx:eqn:assumption}}\\
	&\leq4\beta\|\mD\mU^\top \|_F^2\\
	&\leq 4\beta\left[\frac{1}{8}\|\mX - \mX^\star \|_F^2 + \left(3 + \frac{1}{2\sqrt{2} -2}  \right)\|(\mX - \mX^\star) {\mQ}{\mQ}^\top  \|_F^2\right]. \tag*{By Lemma \ref{nvx:lem:2}}\\
	&\leq4\beta\left[\frac{1}{8}+\left(3 + \frac{1}{2\sqrt{2} -2}\right) \frac{(\beta-\alpha)^2}{(\beta+\alpha)^2} \right]\|\mX-\mX^\star\|_F^2\nn \tag*{By Lemma \ref{nvx:lem:3}}\\
	&\leq 1.76\alpha\|\mX^\star-\mX\|_F^2\tag*{By $ {\beta}/{\alpha}\leq 1.5$}.
	\end{align*}
	
	\paragraph{Combining the two.}
	Hence,
	$$
	\Pi_1+\Pi_2
	\leq-0.24\alpha\|\mX^\star-\mX\|_F^2.
	$$
	Then, we relate the lifted distance $\|\mX^\star-\mX\|_F^2$ with the factored distance $\|\mU-\mU^\star \mR\|_F^2$ using Lemma \ref{nvx:lem:Gongguo} when $r> r^\star$, and Lemma \ref{nvx:lem:ben}  when $r= r^\star$, respectively:
	\begin{align*}
	\text{When $r> r^\star$: }[\nabla^2g(\mU)](\mD,\mD)&\leq-0.24\alpha\min\left\{\rho(\mU)^2,\rho(\mU^\star)^2\right\}\|\mD\|_F^2 \tag*{By Lemma \ref{nvx:lem:Gongguo}}\\
	&=-0.24\alpha\min\left\{\rho(\mU)^2,\rho(\mX^\star)\right\}\|\mD\|_F^2;\\ \\
	\text{When $r= r^\star$: }[\nabla^2g(\mU)](\mD,\mD)&\leq-0.19\alpha\rho(\mU^\star)^2\|\mD\|_F^2\tag*{By Lemma \ref{nvx:lem:ben}}\\
	&=-0.19\alpha\rho(\mX^\star)\|\mD\|_F^2.
	\end{align*}
	For the special case where $\mU=\zero$, we have
	\begin{align*}
	[\nabla^2g(\mU)](\mD,\mD)
	&\leq-0.24\alpha\|\zero-\mX^\star\|_F^2\\
	&=-0.24\alpha\|\mU^\star \mU^{\star \top}\|_F^2\\
	&\leq-0.24\alpha\rho(\mU^\star)^2\|\mU^\star\|_F^2\\
	&=-0.24\alpha\rho(\mX^\star)\|\mD\|_F^2,
	\end{align*}
	where the last second line follows from
	\begin{align*}
	\|\mU^\star \mU^{\star \top}\|_F^2
	=\sum_{i} \sigma_i^4(\mU^\star)
	=\sum_{i:\sigma_i(\mU^\star)\neq0} \sigma_i^4(\mU^\star)
	\geq \min_{i:\sigma_i(\mU^\star)\neq0}\sigma_i^2(\mU^\star) \left(\sum_{j:\sigma_j (\mU^\star)\neq0}\sigma_j^2 (\mU^\star)\right)
	=\rho^2(\mU^\star)\|\mU^\star\|_F^2,
	\end{align*}
	and the last line follows from $\mD=\zero-\mU^\star \mR=-\mU^\star \mR$ when $\mU=\zero.$ Here $\sigma_i(\cdot)$ denotes the $i$-th largest singular value of its argument.
\end{proof}

\section{Understanding the Factored Landscapes for General Non-square Matrices}

In this section, we will study the second convex program~\eqref{nvx:eqn:P1}: the minimization of a general convex function $f(\mX)$ regularized by the matrix nuclear norm $\|\mX\|_*$ with the domain being general matrices. Since the matrix nuclear norm $\|\mX\|_*$ appears in the objective function, the standard convex solvers or even faster tailored ones require performing singular value decomposition in each iteration, which severely limits the efficiency and scalability of the convex program. Motivated by this, we will instead solve its  Burer-Monteiro re-parameterized counterpart.

\subsection{Burer-Monteiro Reformulation of the Nuclear Norm Regularization}

Recall the second problem is the nuclear norm regularization~\eqref{nvx:eqn:P1}:
\begin{align*}
\minimize_{\mX\in\R^{n\times m}} f(\mX)+\lambda\|\mX\|_*\tag{$\mathcal{P}_1$}
\end{align*}
This convex program has an equivalent SDP formulation~\cite[page 8]{nvx:recht2010guaranteed}:
\begin{equation}\label{nvx:eqn:origin:2}
\begin{aligned}
&\minimize_{\mX\in\R^{n\times m},\Phi\in\R^{n\times n},\mPsi\in\R^{m\times m}} f(\mX)+\frac{\lambda}{2}(\tr(\mPhi)+\tr(\mPsi))
~~\st~~ \begin{bmatrix} \mPhi& \mX\\ \mX^\top &\mPsi \end{bmatrix} \succeq 0.
\end{aligned}
\end{equation}
When the  PSD constraint is  implicitly enforced  as the following equality constraint
\begin{align}\label{nvx:eqn:replace:Q}
\begin{bmatrix} \mPhi&\mX\\ \mX^\top &\mPsi \end{bmatrix}=\begin{bmatrix} \mU\\ \mV\end{bmatrix}\begin{bmatrix} \mU\\ \mV\end{bmatrix}^\top
\Rightarrow
\mX=\mU\mV^\top ,
\mPhi=\mU\mU^\top ,
\mPsi=\mV\mV^\top ,
\end{align}
we obtain the Burer-Monteiro factored reformulation~\eqref{nvx:eqn:F1}:
\begin{align*}
\minimize_{\mU\in\R^{n\times r},\mV\in\R^{m\times r}} g(\mU,\mV)= f(\mU\mV^\top )+  \frac{\lambda}{2}(\|\mU\|_F^2+\|\mV\|_F^2)\tag{$\mathcal{F}_1$}.
\end{align*}
The factored formulation~\eqref{nvx:eqn:F1}  can potentially solve the computational issue of~\eqref{nvx:eqn:P1}
in two major respects: {\em (i)} avoiding expensive SVDs by replacing the nuclear norm $\|\mX\|_*$ with the squared term $(\|\mU\|_F^2+\|\mV\|_F^2)/2$; {\em (ii)}  a substantial reduction in the number of the optimization variables from $nm$ to $(n+m)r$.

\subsection{Transforming the Landscape for General Non-square Matrices}

Our primary interest is to understand how the landscape of the lifted objective function $f(\mX)+\lambda\|\mX\|_*$ is transformed by the factored parameterization $\psi(\mU,\mV) = \mU\mV^\top $. The main contribution of this part is establishing that under the restricted well-conditionedness of the convex loss function $f(\mX)$, the factored formulation~\eqref{nvx:eqn:F1} has no spurious local minima and satisfies the strict saddle property.

\begin{theorem}[\bf Transforming the Landscape for General Non-square Matrices]\label{nvx:thm:main:2}
	Suppose the function $f(\mX)$  satisfies the restricted well-conditioned property~\eqref{nvx:eqn:assumption}.
	Assume that $\mX^\star$ of rank $r^\star$ is an optimal solution of ~\eqref{nvx:eqn:P1} where $\lambda>0$. Set $r\geq r^\star$ in the factored program~\eqref{nvx:eqn:F1}.
	Let ${(\mU,\mV)}$ be any critical point of $g(\mU,\mV)$ satisfying $\nabla g(\mU,\mV)=\zero$. Then $(\mU,\mV)$  either corresponds to a  factorization  of  $\mX^\star$, i.e.,
	\begin{align*}
	\mX^\star=\mU\mV^\top ;
	\end{align*}
	or is a strict saddle of the factored problem:
	\begin{align*}
	&\lambda_{\min}(\nabla^2g(\mU,\mV))\leq
	\begin{cases}
	-0.12\alpha\min\left\{0.5\rho^2(\mW),\rho(\mX^\star)\right\}  & \text{ when $r> r^\star$}; \\ \\
	-0.099\alpha\rho(\mX^\star)  & \text{ when $r= r^\star$}; \\ \\
	-0.12\alpha\rho(\mX^\star)  &  \text{ when $\mW= \zero$},
	\end{cases}
	\end{align*}
	where  $W:=\begin{bmatrix} \mU^\top & \mV^\top  \end{bmatrix}^\top $ and $\rho(\mW)$ is the smallest nonzero singular value of $\mW$.
\end{theorem}
Theorem \ref{nvx:thm:main:2} ensures that many local-search algorithms\footnote{The Lipschitz gradient of $g$ at any its sublevel set can be obtained with similar approach for Proposition \ref{nvx:pro:4}.} when applied for solving the factored program~\eqref{nvx:eqn:F1}, can escape from all the saddle points and converge to a global solution that corresponds to $\mX^\star$. Several remarks follow.

\paragraph{The Non-triviality of Extending the PSD Case to the Nonsymmetric Case.}
Although the generalization from the PSD case might not seem technically challenging at first sight, we must overcome several technical difficulties to prove this main theorem. We make a few other technical contributions in the process. In fact, the non-triviality of extending to the nonsymmetric case is also highlighted in~\cite{nvx:tu2015low,nvx:li2016symmetry,nvx:park2017non}. The major technique difficulty to complete such an extension is the ambiguity issue existed in the nonsymmetric case: $\mU\mV^\top =(t\mU)(1/t \mV)^\top $ for any nonzero $t$. This tends to make the factored quadratic objective function badly-conditioned, especially when $t$ is very large or small. To prevent this from happening, a popular strategy utilized to adapt the result for the symmetric case to the non-symmetric case is to introduce an additional balancing regularization to ensure that $\mU$ and $\mV$ have equal energy~\cite{nvx:tu2015low,nvx:li2016symmetry,nvx:park2017non}. Sometimes these additional regularizations are quite complicated (see Eq. (13)-(15) in~\cite{nvx:sun2015guaranteed}). Instead, we find  for nuclear norm regularized problems, the critical points are automatically balanced even without these additional complex balancing regularizations (see Section \ref{nvx:sec:Characterising} for details).   In addition, by connecting the optimality conditions of the convex program~\eqref{nvx:eqn:P1} and the factored program~\eqref{nvx:eqn:F1}, we dramatically simplify the proof argument, making the relationship between the original convex problem and the factored program more transparent.

\paragraph{Proof Sketch of Theorem \ref{nvx:thm:main:2}.}
We try to understand how the parameterization  $\mX= \psi(\mU,\mV)$ transforms the geometric structures of the convex objective function $f(\mX)$  by categorizing  the critical points of the non-convex factored function $g(\mU,\mV)$. In particular, we will illustrate how the globally optimal solution of the convex program is transformed in the domain of $g(\mU,\mV)$.
Furthermore, we will explore the properties of the additional critical points introduced by the parameterization and find a way of utilizing these properties to prove the strict saddle property.  For those purposes, the optimality conditions for the two programs~\eqref{nvx:eqn:P1} and~\eqref{nvx:eqn:F1} will be compared.

\subsection{Optimality Condition for the Convex Program}

As an unconstrained convex optimization, all critical points of~\eqref{nvx:eqn:P1} are global optima and are characterized by the necessary and sufficient KKT condition~\cite{nvx:boyd2004convex}:
\begin{align}\label{nvx:eqn:optimality:origin}
\nabla f(\mX^\star)\in-\lambda\partial\|\mX^\star\|_*,
\end{align}
where $\partial\|\mX^\star\|_*$ denotes the subdifferential (the set of subgradient) of the nuclear norm $\|\mX\|_*$ evaluated at $\mX^\star$.
The subdifferential  of the matrix nuclear norm is defined by
\begin{align*}
\partial \|\mX\|_* =\{&\mD \in \R^{n\times m}:\|\mY\|_* \geq \|\mX\|_*+\langle \mY - \mX, \mD\rangle,
\text{all}\ \mY \in \R^{n\times m}\}.
\end{align*}
We have a more explicit characterization of the subdifferential of the nuclear norm using the singular value decomposition. More specifically, suppose $\mX = \mP\mSigma \mQ^\top $ is the (compact) singular value decomposition of $\mX\in \R^{n \times m}$ with $\mP \in \R^{n\times r}, \mQ \in \R^{m\times r}$ and $\mSigma$ being an $r\times r$ diagonal matrix. Then the subdifferential of the matrix nuclear norm at $\mX$ is given by~\cite[Equation (2.9)]{nvx:recht2010guaranteed}
\begin{align*}
\partial \|\mX\|_* = \{& \mP\mQ^\top  + \mE: \mP^\top  \mE=\zero, \mE\mQ=\zero, \|\mE\| \leq 1\}.
\end{align*}
Combining this representation of the subdifferential and the KKT condition~\eqref{nvx:eqn:optimality:origin} yields an equivalent expression for the optimality condition
\begin{equation}\label{nvx:eqn:optimality:origin:1}
\begin{aligned}
&\nabla f(\mX^\star) \mQ^\star =-\lambda \mP^\star,\\
&\nabla f(\mX^\star)^\top  \mP^\star =-\lambda \mQ^\star,\\
&\|\nabla f(\mX^\star)\|\leq \lambda,
\end{aligned}
\end{equation}
where we assume the compact SVD of  $\mX^\star$ is given by
\[\mX^\star=\mP^\star\mSigma^\star \mQ^{\star \top} \text{ with }\mP^\star \in \R^{n\times r^\star}, \mQ^\star \in \R^{m\times r^\star},  \mSigma^\star\in\R^{r^\star\times r^\star}.  \]
Since   $r\geq r^\star$ in the factored problem~\eqref{nvx:eqn:F1},   to match the dimensions, we define the optimal factors $\mU^\star\in\R^{n\times r}$, $\mV^\star\in\R^{m\times r}$ for any $\mR\in\O_r$ as
\begin{equation}
\begin{aligned}\label{nvx:eqn:global:factor}
\mU^\star&=\mP^\star[\sqrt{\mSigma^\star}~\zero_{r^\star\times(r-r^\star)}] \mR,\\
\mV^\star&=\mQ^\star[\sqrt{\mSigma^\star}~\zero_{r^\star\times(r-r^\star)}] \mR.
\end{aligned}
\end{equation}
Consequently, with the optimal factors $\mU^\star,\mV^\star$ defined in~\eqref{nvx:eqn:global:factor}, we can rewrite the optimal condition~\eqref{nvx:eqn:optimality:origin:1} as
\begin{equation}\label{nvx:eqn:optimality:origin:2}
\begin{aligned}
&\nabla f(\mX^\star) \mV^\star=-\lambda \mU^\star,\\
&\nabla f(\mX^\star)^\top  \mU^\star=-\lambda \mV^\star,\\
&\|\nabla f(\mX^\star)\| \leq \lambda.
\end{aligned}
\end{equation}
Stacking   $\mU^\star,\mV^\star$ as $\mW^\star=\begin{bmatrix} \mU^\star\\ \mV^\star\end{bmatrix}$ and defining
\begin{align}\label{nvx:def:Xi}
\Xi(\mX):= \begin{bmatrix}
\lambda\eye&\nabla f(\mX)\\
\nabla f(\mX)^\top &\lambda\eye
\end{bmatrix}
\text{ for all }\mX
\end{align}
yields a more concise  form of the optimality condition:
\begin{equation}\label{nvx:eqn:optimality:origin:3}
\begin{aligned}
&\Xi(\mX^\star)\mW^\star=\zero,\\
&\|\nabla f(\mX^\star)\| \leq \lambda.
\end{aligned}
\end{equation}

\subsection{Characterizing the Critical Points of the Factored Program}\label{nvx:sec:Characterising}
To begin with, the gradient of $g(\mU,\mV)$ can be computed and rearranged as
\begin{equation}\label{nvx:eqn:grad:1}
\begin{aligned}
\nabla g(\mU,\mV)
&=
\begin{bmatrix}
\nabla_\mU g(\mU,\mV)\\
\nabla_\mV g(\mU,\mV)
\end{bmatrix}
\\
&=
\begin{bmatrix}
\nabla f(\mU\mV^\top)\mV+\lambda \mU\\
\nabla f(\mU\mV^\top)^\top  \mU+\lambda \mV
\end{bmatrix}
\\
&=
\begin{bmatrix}
\lambda\eye&\nabla f(\mU\mV^\top )\\
\nabla f(\mU\mV^\top )^\top &\lambda\eye
\end{bmatrix}
\begin{bmatrix}
\mU\\ \mV
\end{bmatrix}
\\
&=\Xi(\mU\mV^\top )
\begin{bmatrix}
\mU\\ \mV
\end{bmatrix},
\end{aligned}
\end{equation}
where the last equality follows from the definition~\eqref{nvx:def:Xi} of $\Xi(\cdot)$.
Therefore, all critical points of $g(\mU,\mV)$ can be characterized by the following set
\begin{align*}
\calX:= \left\{(\mU,\mV): \Xi(\mU\mV^\top ) \begin{bmatrix} \mU\\ \mV \end{bmatrix}=\zero\right\}.
\end{align*}
We will see that any critical point $(\mU,\mV)\in\calX$ forms an balanced pair, which is defined as follows:
\begin{definition}[\bf Balanced pairs] \label{nvx:def:balanced:pair}
	We call $(\mU,\mV)$ is a balanced pair if the Gram matrices of $\mU$ and $\mV$ are the same:
	$\mU^\top \mU-\mV^\top \mV=\zero.$
	All the balanced pairs form the balanced set, denoted by
	$
	\calE:= \left\{(\mU,\mV): \mU^\top \mU-\mV^\top \mV=\zero\right\}.
	$
\end{definition}

By Definition \ref{nvx:def:balanced:pair}, to show that each critical point forms an balanced pair, we rely on the following fact:
\begin{align}\label{nvx:eqn:balanced:set}
\mW=\begin{bmatrix} \mU\\ \mV  \end{bmatrix},\wh{\mW}=\begin{bmatrix} \mU\\ -\mV  \end{bmatrix} \text{ with }(\mU,\mV)\in\calE \Leftrightarrow  \wh{\mW}^\top {\mW}=W^\top \wh{\mW}=\mU^\top \mU-\mV^\top \mV=\zero.
\end{align}
Now we are ready to relate the critical points and balanced pairs,  the proof of which is given in Appendix \ref{nvx:sec:proof:pro:ef}.
\begin{proposition}\label{nvx:pro:ef}
	Any critical point $(\mU,\mV)\in\calX$ forms a balanced pair in $\calE.$
\end{proposition}
\subsubsection{The Properties of the Balanced Set}
In this part, we introduce some important properties of the balanced set $\calE$. These properties basically compare   the on-diagonal-block energy and the off-diagonal-block energy for a certain block matrix. Hence, it is necessary to introduce two operators defined on block matrices:
\begin{equation}\label{nvx:def:Pon:Poff}
\begin{aligned}
\Pon\left(\begin{bmatrix} \mA_{11} &\mA_{12}\\ \mA_{21}&\mA_{22}  \end{bmatrix}\right)&:=\begin{bmatrix} \mA_{11}&\zero  \\ \zero& \mA_{22}  \end{bmatrix}, \\
\Poff\left(\begin{bmatrix} \mA_{11} &\mA_{12}\\ \mA_{21}&\mA_{22}  \end{bmatrix}\right)&:=\begin{bmatrix} \zero&A_{12} \\ \mA_{21}&\zero  \end{bmatrix},
\end{aligned}
\end{equation}
for any matrices $\mA_{11}\in\R^{n\times n}, \mA_{12}\in\R^{n\times m}, \mA_{21}\in\R^{m\times n}, \mA_{22}\in\R^{m\times m}$.

According to the definitions of $\Pon$ and $\Poff$ in~\eqref{nvx:def:Pon:Poff},
when $\Pon$ and $\Poff$ are acting on the product of two block matrices $\mW_1\mW_2^\top $,
\begin{equation}\label{nvx:eqn:Pon:Poff}
\begin{aligned}
\Pon(\mW_1\mW_2^\top )&=\Pon\left(\begin{bmatrix}\mU_1\mU_2^\top  & \mU_1\mV_2^\top \\ \mV_1\mU_2^\top &\mV_1\mV_2^\top \end{bmatrix}\right)=\begin{bmatrix} \mU_1\mU_2^\top &\zero  \\ \zero& \mV_1\mV_2^\top   \end{bmatrix}= \frac{\mW_1\mW_2^\top +\wh{\mW}_1\wh{\mW}_2^\top }{2},\\
\Poff(\mW_1\mW_2^\top )&=\Pon\left(\begin{bmatrix}\mU_1\mU_2^\top  & \mU_1\mV_2^\top \\\mV_1\mU_2^\top &\mV_1\mV_2^\top \end{bmatrix}\right)=\begin{bmatrix} \zero&\mV_1\mV_2^\top   \\ \mV_1\mU_2^\top & \zero \end{bmatrix}= \frac{\mW_1\mW_2^\top -\wh{\mW}_1\wh{\mW}_2^\top }{2}.
\end{aligned}
\end{equation}
Here, to simplify the notations, for any $\mU_1,\mU_2 \in\R^{n\times r}$ and $\mV_1,\mV_2\in\R^{m\times r}$, we define
\[\mW_1=\begin{bmatrix} \mU_1\\ \mV_1 \end{bmatrix},\qquad \wh{\mW}_1=\begin{bmatrix} \mU_1\\-\mV_1 \end{bmatrix},\qquad \mW_2=\begin{bmatrix} \mU_2\\ \mV_2 \end{bmatrix},\qquad \wh{\mW}_2=\begin{bmatrix} \mU_2\\-\mV_2 \end{bmatrix}.\]

Now, we are ready to present the properties regarding the set $\calE$ in Lemma \ref{nvx:lem:ef:1} and  Lemma \ref{nvx:lem:ef:2}, whose proofs are given in Appendix \ref{nvx:sec:proof:lem:ef:1} and  Appendix \ref{nvx:sec:proof:lem:ef:2}, respectively.
\begin{lemma}\label{nvx:lem:ef:1} Let $\mW=\begin{bmatrix} \mU^\top &\mV^\top  \end{bmatrix}^\top$  with $(\mU,\mV)\in\calE$. Then for every $\mD=\begin{bmatrix}\mD_\mU^\top &\mD_\mV^\top  \end{bmatrix}^\top$ of proper dimension,
	we have
	\[\|\Pon(\mD\mW^\top )\|_F^2=\|\Poff(\mD\mW^\top )\|_F^2.\]
	
\end{lemma}

\begin{lemma}\label{nvx:lem:ef:2}
	Let $\mW_1=\begin{bmatrix} \mU_1^\top &\mV_1^\top  \end{bmatrix}^\top$, $\mW_2=\begin{bmatrix} \mU_2^\top &\mV_2^\top  \end{bmatrix}^\top$ with $(\mU_1,\mV_1),(\mU_2,\mV_2)\in\calE$. Then
	\[\|\Pon(\mW_1\mW_1^\top -\mW_2 \mW_2^\top )\|_F^2\leq\|\Poff(\mW_1\mW_1^\top -\mW_2 \mW_2^\top )\|_F^2. \]
\end{lemma}

\subsection{Proof Idea: Connecting the Optimality Conditions}

First observe that each $(\mU^\star,\mV^\star)$   in~\eqref{nvx:eqn:global:factor}  is a global optimum for the factored program (we prove this in Appendix \ref{nvx:sec:proof:pro:global}):
\begin{proposition}\label{nvx:pro:global}
	Any  $(\mU^\star,\mV^\star)$   in~\eqref{nvx:eqn:global:factor} is  a global optimum of the factored program~\eqref{nvx:eqn:F1}:
	\[g(\mU^\star,\mV^\star)\leq g(\mU,\mV),\text{ for all }\mU\in\R^{n\times r}, \mV\in\R^{m\times r}.\]
\end{proposition}

However, due to non-convexity,   only characterizing the global optima is not enough for the factored program to achieve the global convergence by many local-search algorithms. One should also eliminate the possibility of the existence of spurious local minima or degenerate saddles.
For this purpose, we focus on the critical point set $\calX$ and  observe that   any critical point $(\mU,\mV)\in\calX$ of the factored problem satisfies the first part of the optimality condition~\eqref{nvx:eqn:optimality:origin:3}:
$$\Xi(\mX)\mW=\zero$$
by constructing $\mW=[\mU^\top ~\mV^\top ]^\top $ and $\mX= \mU\mV^\top $. If the critical point $(\mU,\mV)$ additionally satisfies $\|\nabla f(\mU\mV^\top )\|\leq\lambda$,  then it corresponds to the global optimum $\mX^\star=\mU\mV^\top $.

Therefore, it remains to study the additional critical points (which are introduced by the parameterization $\mX= \psi(\mU,\mV)$) that violate $\|\nabla f(\mU\mV^\top )\|\leq\lambda$.
In fact,
we intend to show the following: for any critical point $(\mU,\mV)$, if $\mX^\star\neq \mU\mV^\top $, we can find a direction $\mD$, in which the Hessian $\nabla^2 g(\mU,\mV)$ has a strictly negative curvature $[\nabla^2 g(\mU,\mV)](\mD,\mD)<-\tau \|\mD\|_F^2$ for some $\tau>0$. Hence, every critical point $(\mU,\mV)$ either corresponds to the global optimum $\mX^\star$, or is a strict saddle point.

To gain more intuition,  we take a closer look at the directional curvature of $g(\mU,\mV)$ in some direction $\mD=[\mD_\mU^\top ~\mD_\mV^\top ]^\top $:
\begin{equation}\label{nvx:eqn:Hessian}
\begin{aligned}
&[\nabla^2 g(\mU,\mV)](\mD,\mD)=
\lg\Xi(\mX)
, \mD\mD^\top \rg+
[\nabla^2 f(\mX)](\mD_\mU\mV^\top +\mU\mD_\mV^\top ,\mD_\mU\mV^\top +\mU\mD_\mV^\top ),
\end{aligned}
\end{equation}
where the second term is always nonnegative by the convexity of $f$. The sign of the first term $\lg\Xi(\mX), \mD\mD^\top \rg$  depends on the positive semi-definiteness of $\Xi(\mX)$, which is related to the boundedness condition $\|\nabla f(\mX)\|\leq \lambda$ through the Schur complement theorem~\cite[A.5.5]{nvx:boyd2004convex}:
\begin{align*}
\Xi(\mX)\succeq 0 \Leftrightarrow \lambda \eye-\frac{1}{\lambda}\nabla f(\mX)^\top  \nabla f(\mX)\succeq 0
\Leftrightarrow \|\nabla f(\mX)\|\leq\lambda.
\end{align*}
Equivalently, whenever $\|\nabla f(\mX)\|>\lambda$, we have $\Xi(\mX)\nsucceq 0$. Therefore, for those non-globally optimal critical points $(\mU,\mV)$, it is possible to
find a direction $\mD$ such that the first term $\lg\Xi(\mX), \mD\mD^\top \rg$ is strictly negative. Inspired by the weighted PCA example, we choose  $\mD$
as the direction from the critical point $\mW=\begin{bmatrix} \mU^\top &\mV^\top  \end{bmatrix}^\top$ to the nearest globally optimal  factor $\mW^\star \mR$ with $\mW^\star=\begin{bmatrix} {\mU^\star}^\top &{\mV^\star}^\top \end{bmatrix}^\top$, i.e.,
\[\mD=\mW-\mW^\star \mR,\]
where $\mR=\argmin_{\mR:\mR\mR^\top =\eye_r}\|\mW-\mW^\star \mR\|_F$.
We will see that with this particular $\mD$,  the first term of~\eqref{nvx:eqn:Hessian}  will be strictly negative while the second term retains small.

\subsection{A Formal Proof of Theorem \ref{nvx:thm:main:2}}

The main argument involves choosing $\mD$ as the direction from $\mW=\begin{bmatrix} \mU^\top &\mV^\top  \end{bmatrix}^\top$ to its nearest optimal factor: $\mD=\mW-\mW^\star \mR $ with $\mR=\argmin_{\mR:\mR\mR^\top =\eye_r}\|\mW-\mW^\star \mR\|_F$, and  showing that the Hessian $\nabla^2 g(\mU,\mV)$ has a strictly negative curvature in the direction of $\mD$ whenever $\mW\neq \mW^\star$.
To that end, we  first introduce the following lemma (with its proof in Appendix \ref{nvx:app:lem:bound:QQ}) connecting the distance $\|\mU\mV^\top -\mX^\star\|_F$ and the distance $\|(\mW\mW^\top -\mW^\star \mW^{\star \top})\mQ\mQ^\top  \|_F$ (where $\mQ\mQ^\top $ is an orthogonal projector onto the $\tmop{Span}(\mW)$).
\begin{lemma}\label{nvx:lem:bound:QQ}
	Suppose the function $f(\mX)$ in \eqref{nvx:eqn:P1} is restricted well-conditioned~\eqref{nvx:eqn:assumption}.
	Let $\mW=\begin{bmatrix} \mU^\top &\mV^\top  \end{bmatrix}^\top$ with $(\mU,\mV)\in\calX$,
$\mW^\star=\begin{bmatrix} {\mU^\star}^\top &{\mV^\star}^\top \end{bmatrix}^\top$ correspond to the global optimum of~\eqref{nvx:eqn:P1}
	and  $\mQ\mQ^\top $ be the orthogonal projector onto $\range(\mW)$. Then
	\begin{align*}
	\|(\mW\mW^\top -\mW^\star \mW^{\star \top})\mQ\mQ^\top  \|_F \leq  2\frac{\beta-\alpha}{\beta+\alpha}\|\mU\mV^\top -\mX^\star\|_F.
	\end{align*}
\end{lemma}

\begin{proof}[\bf Proof of Theorem  \ref{nvx:thm:main:2}]
	Let $\mD=W-\mW^\star \mR$ with $\mR=\argmin_{\mR:\mR\mR^\top =\eye_r}\|W-\mW^\star \mR\|_F$.
	Then
	\begin{align*}
	&[\nabla^2 g(\mU,\mV)](\mD,\mD)\\
	&=
	\lg\Xi(\mX), \mD\mD^\top \rg+[\nabla^2 f(\mX)](\mD_\mU\mV^\top +\mU\mD_\mV^\top ,\mD_\mU\mV^\top +\mU\mD_\mV^\top )\\
	&\stack{\ding{172}}{=}
	\lg\Xi(\mX)
	, \mW^\star \mW^{\star \top}-\mW\mW^\top \rg+[\nabla^2 f(\mX)](\mD_\mU\mV^\top +\mU\mD_\mV^\top ,\mD_\mU\mV^\top +\mU\mD_\mV^\top )\\
	&\stack{\ding{173}}{\leq}
	\left\lg\Xi(\mX)-\Xi(\mX^\star)
	, \mW^\star \mW^{\star \top}-\mW\mW^\top \right\rg+[\nabla^2 f(\mX)](\mD_\mU\mV^\top +\mU\mD_\mV^\top ,\mD_\mU\mV^\top +\mU\mD_\mV^\top )\\
	&=
	\left\lg
	\begin{bmatrix}
	\lambda\eye&\nabla f(\mX)\\
	\nabla f(\mX)^\top &\lambda\eye
	\end{bmatrix}-
	\begin{bmatrix}
	\lambda\eye&\nabla f(\mX^\star)\\
	\nabla f(\mX^\star)^\top &\lambda\eye
	\end{bmatrix}
	, \mW^\star \mW^{\star \top}-\mW\mW^\top \right\rg
	+[\nabla^2 f(\mX)](\mD_\mU\mV^\top +\mU\mD_\mV^\top ,\mD_\mU\mV^\top +\mU\mD_\mV^\top )
	\\
	&\stack{\ding{174}}{=}
	\left\lg\begin{bmatrix}
	\zero&\int_{0}^1[\nabla^2 f(\mX^\star+t(\mX-\mX^\star))](\mX-\mX^\star)d t\\
	*&\zero
	\end{bmatrix}
	, \mW^\star \mW^{\star \top}-\mW\mW^\top \right\rg
	+[\nabla^2 f(\mX)](\mD_\mU\mV^\top +\mU\mD_\mV^\top ,\mD_\mU\mV^\top +\mU\mD_\mV^\top )\\
	&=
	-2\int_{0}^1[\nabla^2 f(\mX^\star+t(\mX-\mX^\star))](\mX-\mX^\star,\mX-\mX^\star)d t
	+[\nabla^2 f(\mX)](\mD_\mU\mV^\top +\mU\mD_\mV^\top ,\mD_\mU\mV^\top +\mU\mD_\mV^\top )
	\end{align*}
	where \ding{172} follows from $\nabla g(\mU,\mV)=\Xi(\mX)\mW=\zero$ and~\eqref{nvx:eqn:grad:1}. For \ding{173}, we note that
	$\lg \Xi(\mX^\star), \mW^\star \mW^{\star \top}-\mW\mW^\top \rg\leq0$
	since $ \Xi(\mX^\star) \mW^\star=\zero$ in~\eqref{nvx:eqn:optimality:origin:3} and $  \Xi(\mX^\star)\succeq 0$ by the optimality condition.  For \ding{174}, we first use $*=(\int_{0}^1[\nabla^2 f(\mX^\star+t(\mX-\mX^\star))](\mX-\mX^\star)d t)^\top $ for convenience and then it
	follows from the Taylor's Theorem for vector-valued functions ~\cite[Eq. (2.5) in Theorem 2.1]{nvx:nocedal2006numerical}:
	\[\nabla f(\mX)-\nabla f(\mX^\star)=\int_{0}^1[\nabla^2 f(\mX^\star+t(\mX-\mX^\star))](\mX-\mX^\star)d t.\]
	
	Now, we continue the argument:
    {\allowdisplaybreaks
	\begin{align*}
	&[\nabla^2 g(\mU,\mV)](\mD,\mD)\\
	&\leq
	-2\int_{0}^1[\nabla^2 f(\mX^\star+t(\mX-\mX^\star))](\mX-\mX^\star,\mX-\mX^\star)d t
	+[\nabla^2 f(\mX)](\mD_\mU\mV^\top +\mU\mD_\mV^\top ,\mD_\mU\mV^\top +\mU\mD_\mV^\top )\\
	&\stack{\ding{175}}{\leq} -2\alpha\|\mX^\star-\mX\|_F^2+\beta\|\mD_\mU\mV^\top +\mU\mD_\mV^\top \|_F^2,\\
	&\stack{\ding{176}}{\leq} -0.5\alpha \|\mW\mW^\top -\mW^\star \mW^{\star \top}\|_F^2+2\beta(\|\mD_\mU\mV^\top \|_F^2+\|\mU\mD_\mV^\top \|_F^2)\\
	&\stack{\ding{177}}{=} -0.5\alpha \|\mW\mW^\top -\mW^\star \mW^{\star \top}\|_F^2+\beta\|\mD\mW^\top \|_F^2\\
	&\stack{\ding{178}}{\leq}  \left[-0.5{\alpha}+{\beta}/{8}+ 4.208 \beta \left(\frac{\beta-\alpha}{\beta+\alpha}\right)^2 \right]
	\|\mW\mW^\top -\mW^\star \mW^{\star \top}\|_F^2
	\\
	&\stack{\ding{179}}{\leq} -0.06\alpha\|\mW\mW^\top -\mW^\star \mW^{\star \top}\|_F^2\\
	&\stack{\ding{180}}{\leq}
	\begin{cases}
	-0.06\alpha\min\left\{\rho^2(\mW),\rho^2(\mW^\star)\right\}\|\mD\|_F^2,  &\qquad\qquad\text{By Lemma \ref{nvx:lem:Gongguo} when $r>r^\star$}
	\\  \\
	-0.0495\alpha\rho^2(\mW^\star)\|\mD\|_F^2,  &\qquad\qquad\text{By Lemma \ref{nvx:lem:ben} when $r=r^\star$}
	\\  \\
	-0.06\alpha\rho^2(\mW^\star)\|\mD\|_F^2,  &\qquad\qquad\text{When $\mW=\zero$}
	\end{cases}
	\end{align*}
    }
	where \ding{175} uses the restricted well-conditionedness  ~\eqref{nvx:eqn:assumption} since $\rank(\mX^\star+t(\mX-\mX^\star))\leq 2r$, $\rank(\mX-\mX^\star)\leq 4r$ and $\rank(\mD_\mU\mV^\top +\mU\mD_\mV^\top )\leq 4r.$
	\ding{176} comes from Lemma \ref{nvx:lem:ef:2} and the fact $\|\mA+\mB\|_F^2\leq 2(\|\mA\|_F^2+\|\mB\|_F^2)$. \ding{177} follows from Lemma \ref{nvx:lem:ef:1}.
	\ding{178}  first  uses  Lemma \ref{nvx:lem:2} to bound $\|\mD\mW^\top \|_F^2=\|(\mW-\mW^\star \mR)\mW^\top \|_F^2$ since $\mW^\top \mW^\star\succeq0$ and then uses Lemma \ref{nvx:lem:bound:QQ} to further bound $\|(\mW^\star-\mW)\mQ\mQ^\top \|_F^2$.
	\ding{179} holds when $\beta/\alpha\leq1.5$.
	\ding{180}  uses the similar argument as in the proof of Theorem \ref{nvx:thm:main:1} to relate the lifted distance and factored distance.
	Particularly,  three possible cases are considered:
	(i) $r>r^\star$; (ii) $r=r^\star$; (iii) $\mW=\zero$.
	We apply Lemma \ref{nvx:lem:Gongguo} to Case (i) and Lemma  \ref{nvx:lem:ben} to Case (ii).
	For the third case that $\mW=\zero$,  we obtain from \ding{179} that
	\begin{align*}
	[\nabla^2 g(\mU,\mV)](\mD,\mD)
	&\leq-0.06\alpha \|\mW^\star \mW^{\star \top}\|_F^2
	\leq
	-0.06\alpha \rho(\mW^\star)^2\|\mW^\star\|_F^2
	=-0.06\alpha \rho(\mW^\star)^2\|\mD\|_F^2,
	\end{align*}
	where the last equality follows from $\mD=\zero-\mW^\star \mR=-\mW^\star \mR$ because $\mW=\zero.$

	The final result follows from the the definition of $\mU^\star,\mV^\star$ in~\eqref{nvx:eqn:global:factor}:
	$$
	\mW^\star = \begin{bmatrix} \mP^\star\sqrt{\mSigma^\star}\mR \\ \mQ^\star\sqrt{\mSigma^\star}\mR \end{bmatrix} = \begin{bmatrix}\mP^\star/\sqrt{2} \\ \mQ^\star/\sqrt{2} \end{bmatrix} \left(\sqrt{2\mSigma^\star}\right)\mR,
	$$
	which implies $\sigma_\ell(\mW^\star)=\sqrt{2\sigma_\ell(\mX^\star)}.$
\end{proof}

\section{Conclusion}
In this work, we considered two popular  minimization problems:  the minimization of a general convex function $f(\mX)$ with the domain being positive semi-definite matrices;  the minimization of a general convex function $f(\mX)$ regularized by the matrix nuclear norm $\|\mX\|_*$ with the domain being general matrices. To improve the computational efficiency,  we applied the Burer-Monteiro re-parameterization and showed that,
as long as the convex function $f(\mX)$ is (restricted) well-conditioned,   the resulting factored problems have the following properties: each critical point either corresponds to a global optimum of the original convex programs, or is a strict saddle where the Hessian matrix has a strictly negative eigenvalue. Such a benign landscape then allows many iterative optimization methods to escape from all the saddle points and converge to a global optimum with even random initializations.

\begin{appendices}
		\section{Proof of Proposition \ref{nvx:pro:4}}\label{nvx:sec:proof:pro:4}
		
		To that end, we first show that for any $\mU\in\tmop{Lev}_f({\mU_0})$, $\|\mU\|_F $ is upper-bounded.
		Let $\mX=\mU\mU^\top$ and consider the following second-order Taylor expansion of $f(\mX)$
		\begin{align*}
		f(\mX)&=f(\mX^\star)+\lg \nabla f(\mX^\star), \mX-\mX^\star\rg+ \frac{1}{2}\int_0^1[\nabla^2 f(t \mX^\star+ (1-t)\mX)](\mX-\mX^\star,\mX-\mX^\star)dt\\
		&\geq f(\mX^\star) + \frac{1}{2}\int_0^1[\nabla^2 f(t \mX^\star+ (1-t)\mX)](\mX-\mX^\star,\mX-\mX^\star)dt\\
		& \geq f(\mX^\star) + \frac{\alpha}{2}\|\mX-\mX^\star\|_F^2,
		\end{align*}
		which implies that
		\begin{align}\label{nvx:eqn:lip1}
		\|\mU\mU^\top-\mX^\star\|_F^2\leq \frac{2}{\alpha}(f(\mU\mU^\top) - f(\mX^\star)) \leq \frac{2}{\alpha}(f(\mU_0\mU_0^\top) - f(\mX^\star))
		\end{align}
		with the second inequality following from the assumption $U\in\tmop{Lev}_f({\mU_0})$.
		Thus, we have
		\begin{align}\label{nvx:eqn:lip2}
		\|\mU\|_F \leq \|\mU^\star\|_F + \dist(\mU,\mU^\star)\leq \|\mU^\star\|_F + \frac{\|\mU\mU^\top-\mX^\star\|_F}{2(\sqrt{2} -1)\rho(\mU^\star)}
		\leq \|\mU^\star\|_F + \frac{\sqrt{\frac{2}{\alpha}(f(\mU_0\mU_0^\top) - f(\mX^\star))}}{2(\sqrt{2} -1)\rho(\mU^\star)}.
		\end{align}
		Now we are ready to show the Lipschitz gradient for $g$ at $\tmop{Lev}_f({\mU_0})$:
		\begin{align*}
		\|\nabla^2 g(\mU)\|^2 &= \max_{\|\mD\|_F=1}\left|[\nabla^2g(\mU)](\mD,\mD)\right|\\
		&= \max_{\|\mD\|_F=1} \left|2\lg\nabla f(\mU\mU^\top),\mD\mD^\top \rg +[\nabla^2f(\mU\mU^\top)](\mD\mU^\top +\mU\mD^\top ,\mD\mU^\top  +\mU\mD^\top )\right|\\
		&\leq 2\max_{\|\mD\|_F=1} \left|\lg\nabla f(\mU\mU^\top),\mD\mD^\top \rg\right| + \max_{\|\mD\|_F=1} \left|[\nabla^2f(\mU\mU^\top)](\mD\mU^\top +\mU\mD^\top ,\mD\mU^\top  +\mU\mD^\top)\right| \\
		& \leq 2\max_{\|\mD\|_F=1} \left|\lg\nabla f(\mU\mU^\top) - \nabla f(\mX^\star),\mD\mD^\top \rg\right| + 2\| \nabla f(\mX^\star) \|_F + \beta \| \mD\mU^\top +\mU\mD^\top \|_F^2\\
		& \leq 2 \beta\|\mU\mU^\top - \mX^\star\|_F + 2\| \nabla f(\mX^\star) \|_F + 4\beta \|\mU\|_F^2\\
		&\leq 2\beta \sqrt{\frac{2}{\alpha}(f(\mU_0\mU_0^\top) - f(\mX^\star))} + 2\| \nabla f(\mX^\star) \|_F + 4\beta \left(\|\mU^\star\|_F + \frac{\sqrt{\frac{2}{\alpha}(f(\mU_0\mU_0^\top) - f(\mX^\star))}}{2(\sqrt{2} -1)\rho(\mU^\star)}\right)^2\\
		&:=L_c^2.
		\end{align*}
		Here, the last second line follows from  ~\eqref{nvx:eqn:lip1} and~\eqref{nvx:eqn:lip2}.
		This concludes the proof of  Proposition \ref{nvx:pro:4}.
		\hfill$\square$

	\section{Proof of Lemma \ref{nvx:lem:Gongguo}}\label{nvx:sec:proof:lem:Gongguo}
	
	Let $\mX_1=\mU_1\mU_1^\top $,  $\mX_2=\mU_2\mU_2^\top $ and their full eigenvalue decompositions be
	\begin{align*}
	\mX_1&=\sum_{j=1}^n\lambda_j\p_j\p_j^\top , \qquad
	\mX_2=\sum_{j=1}^n\eta_j\q_j\q_j^\top
	\end{align*}
	where $\{\lambda_j\}$ and $\{\eta_j\}$ are the eigenvalues in decreasing order. Since $\rank(\mU_1) = r_1$ and $\rank(\mU_2) = r_2$, we have $\lambda_{j}=0$ for $j> r_1$ and $\eta_j=0$ for $j> r_2$.
	We compute $\|\mX_1-\mX_2\|_F^2$ as follows
    {\allowdisplaybreaks
	\begin{align*}
	\|\mX_1-\mX_2\|_F^2
	&=\|\mX_1\|_F^2+\|\mX_2\|_F^2-2\lg \mX_1,\mX_2\rg\\
	&=\sum_{i=1}^n\lambda_i^2+\sum_{j=1}^n\eta_j^2 - \sum_{i=1}^n\sum_{j=1}^n2\lambda_i\eta_j\lg \p_i,\q_j\rg^2\\
	&\stack{\ding{172}}{=}\sum_{i=1}^n\lambda_i^2\sum_{j=1}^n\lg \p_i,\q_j\rg^2+\sum_{j=1}^n\eta_j^2\sum_{i=1}^n\lg \p_i,\q_j\rg^2 - \sum_{i=1}^n\sum_{j=1}^n2\lambda_i\eta_j\lg \p_i,\q_j\rg^2\\
	&\stack{\ding{173}}{=}\sum_{i=1}^{ n}\sum_{j=1}^{n}(\lambda_i-\eta_j)^2\lg \p_i,\q_j\rg^2\\
	& {~}{=}\sum_{i=1}^{ n}\sum_{j=1}^{ n}\left(\sqrt{\lambda_i}-\sqrt{\eta_j}\right)^2\left(\sqrt{\lambda_i}+\sqrt{\eta_j}\right)^2\lg \p_i,\q_j\rg^2\\
	& \stack{\ding{174}}{\geq}\min\left\{ \sqrt{\lambda_{ r_1}}, \sqrt{\eta_{r_2}}\right\}^2\sum_{i=1}^{ n}\sum_{j=1}^{ n}\left(\sqrt{\lambda_i}-\sqrt{\eta_j}\right)^2\lg \p_i,\q_j\rg^2\\
	& \stack{\ding{175}}{=}\min\left\{ {\lambda_{ r_1}}, {\eta_{r_2}}\right\}\left\| \sqrt{\mX_1}- \sqrt{\mX_2}\right\|_F^2,
	\end{align*}
    }
	where \ding{172} uses the fact $\sum_{j=1}^n\lg \p_i,\q_j\rg^2=\|\p_i\|_2^2=1$ with $\{\q_j\}
	$ being an orthonormal basis and similarly $\sum_{i=1}^n\lg \p_i,\q_j\rg^2$ $=\|\q_j\|_2^2=1$.
	\ding{173} is by firstly an exchange of the summations, secondly the fact that $\lambda_{j}=0$ for $j> r_1$ and $\eta_j=0$ for $j> r_2$, and thirdly completing squares.
	\ding{174} is because $\{\lambda_j\}$ and $\{\eta_j\}$ are sorted in decreasing order.
	\ding{175} follows from  \ding{173} and that $\{\sqrt{\lambda_j}\}$ and  $\{\sqrt{\eta_j}\}$ are eigenvalues of $\sqrt{\mX_1}$ and $\sqrt{\mX_2}$, the matrix square root of $\mX_1$ and $\mX_2$, respectively.
	
	Finally, we can conclude the proof as long as we can show the following inequality:
	\begin{align}\label{nvx:eqn:sqrtX}
	\left\|\sqrt{\mX_1}-\sqrt{\mX_2}\right\|_F^2\geq \min_{\mR: \mR\mR^\top =\eye_r}\|\mU_1-\mU_2\mR\|_F^2.
	\end{align}
	By expanding $\|\cdot\|_F^2$ in \eqref{nvx:eqn:sqrtX} and noting that $\lg \sqrt{\mX_1},\sqrt{\mX_1}\rg=\tr(\mX_1)=\tr(\mU_1\mU_1^\top )$ and $\lg \sqrt{\mX_2},\sqrt{\mX_2}\rg=\tr(\mX_2)=\tr(\mU_2\mU_2^\top )$,~\eqref{nvx:eqn:sqrtX} reduces to
	\begin{align}\label{nvx:eqn:sqrtX:2}
	\lg \sqrt \mX_1,\sqrt{\mX_2}\rg\leq \max_{\mR: \mR\mR^\top =\eye_r}\lg \mU_1, \mU_2\mR\rg.
	\end{align}
	To show~\eqref{nvx:eqn:sqrtX:2}, we write the SVDs  of $\mU_1, \mU_2$ respectively as $\mU_1=\mP_1\mSigma_1\mQ_1^\top $ and $\mU_2=\mP_2\mSigma_2\mQ_2^\top $ with $\mP_1, \mP_2\in\R^{n\times r}$, $\mSigma_1,\mSigma_2\in\R^{r\times r}$ and
	$\mQ_1,\mQ_2\in\R^{r\times r}$. Then we have $\sqrt{\mX_1}=\mP_1\mSigma_1\mP_1^\top ,\sqrt{\mX_2}=\mP_2\mSigma_2\mP_2^\top .$
	
	On one hand,
	\begin{align*}
	\text{RHS of~\eqref{nvx:eqn:sqrtX:2}}
	&=\max_{\mR: \mR\mR^\top =\eye_r}\left\lg \mP_1\mSigma_1\mQ_1^\top , \mP_2\mSigma_2\mQ_2^\top  \mR\right\rg\\
	&=\max_{\mR: \mR\mR^\top =\eye_r}\left\lg \mP_1\mSigma_1,\mP_2\mSigma_2,\mQ_2^\top  \mR  \mQ_1  \right\rg\\
	&= \max_{\mR: \mR\mR^\top =\eye_r}\left\lg \mP_1\mSigma_1,\mP_2\mSigma_2 \mR           \right\rg\tag*{By  $\mR\leftarrow  \mQ_2^\top  \mR  \mQ_1$}\\
	&= \| (\mP_2\mSigma_2)^\top \mP_1\mSigma_1\|_*. \tag*{By Lemma \ref{nvx:lem:procrust}}
	\end{align*}
	
	On the other hand,
	\begin{align*}
	\text{LHS of~\eqref{nvx:eqn:sqrtX:2}}&=\lg \mP_1\mSigma_1\mP_1^\top ,  \mP_2\mSigma_2\mP_2^\top \rg\\
	&=\lg (\mP_2\mSigma_2)^\top \mP_1\mSigma_1,\mP_2^\top  \mP_1\rg\\
	&\leq \| (\mP_2\mSigma_2)^\top \mP_1\mSigma_1\|_*\|\mP_2^\top  \mP_1\| \tag*{By H\"{o}lder's Inequality}\\
	&\leq \| (\mP_2\mSigma_2)^\top \mP_1\mSigma_1\|_* \tag*{Since  $\|\mP_2^\top  \mP_1\|\leq \|\mP_2\|\|\mP_1\|\leq 1$}.
	\end{align*}
	This proves~\eqref{nvx:eqn:sqrtX:2} and hence completes the proof of Lemma \ref{nvx:lem:Gongguo}.
	\hfill$\square$

	\section{Proof of Lemma \ref{nvx:lem:2}}\label{nvx:sec:proof:lem2}
	
	The proof relies on the following lemma.
	\begin{lemma}\label{nvx:lem:1}\cite[Lemma E.1]{nvx:bhojanapalli2016lowrankrecoveryl}
		Let $\mU$ and $ \mZ $ be any two  matrices in $\R^{n\times r}$ such that $\mU^\top    \mZ  =  \mZ^\top   \mU$ is PSD.
		Then
		\[
		\left\|\left(\mU - \mZ  \right)\mU^\top  \right\|_F^2\leq
		\frac{1}{2\sqrt{2} -2}\left\|\mU\mU^\top   -  \mZ  \mZ^\top     \right\|_F^2.
		\]
	\end{lemma}

	\begin{proof}[\bf Proof of Lemma \ref{nvx:lem:2}]
		Define two orthogonal projectors
		\[\calQ=\mQ\mQ^\top  \qquad\text{and}\qquad\calQ_\bot=\mQ_{\bot}\mQ_{\bot}^\top ,\]
		so $\calQ$ is the orthogonal projector onto  $\range(\mU)$ and  $\calQ_\bot$ is the orthogonal projector onto the orthogonal complement of $\range(\mU)$.  Then
		\begin{align}
		\|(\mU- \mZ )\mU^\top  \|_F^2
		&\stack{\ding{172}}{=}\|(\mU -\calQ  \mZ ) \mU^\top  \|_F^2+\|\calQ_\bot \mU^\top  \|_F^2\nn\\
		&\stack{\ding{173}}{=}\|(\mU -\calQ  \mZ ) \mU^\top  \|_F^2+\lg { \mZ^\top }\calQ_\bot  \mZ ,\mU^\top   \mU\rg\nn\\
		&\stack{\ding{174}}{\leq}\frac{1}{2\sqrt2-2}\|\mU\mU^\top  -(\calQ  \mZ) (\calQ  \mZ)^\top  \|_F^2+\lg { \mZ^\top }\calQ_\bot  \mZ ,\mU^\top   \mU-{ \mZ^\top }\calQ \mZ \rg+\lg { \mZ^\top }\calQ_\bot  \mZ ,{ \mZ^\top }\calQ \mZ \rg\nn\\
		&\stack{\ding{175}}{\leq}\frac{1}{2\sqrt2-2}\|\mU\mU^\top  -\calQ  \mZ   \mZ^\top  \|_F^2+\lg { \mZ^\top }\calQ_\bot  \mZ ,\mU^\top   \mU-{ \mZ^\top }\calQ \mZ \rg+\lg { \mZ^\top }\calQ_\bot  \mZ ,{ \mZ^\top }\calQ \mZ \rg\nn\\
		&\stack{\ding{176}}{\leq}\frac{1}{2\sqrt2-2}\|\mU\mU^\top  -\calQ  \mZ   \mZ^\top  \|_F^2+\frac{1}{8}\|{ \mZ^\top }\calQ_\bot  \mZ \|_F^2+2\|\mU^\top   \mU-{ \mZ^\top }\calQ  \mZ \|_F^2+\lg { \mZ^\top }\calQ_\bot  \mZ ,{ \mZ^\top }\calQ  \mZ \rg, \label{nvx:eqn:final:step}
		\end{align}
		where \ding{172} is by expressing $(\mU- \mZ )\mU^\top $ as the sum of two orthogonal factors $(\mU -\calQ  \mZ ) \mU^\top $ and  $-\calQ_\bot  \mZ  \mU^\top $.
		\ding{173} is because $\|\calQ_\bot \mZ  \mU^\top  \|_F^2=\lg\calQ_\bot \mZ  \mU^\top  ,\calQ_\bot \mZ  \mU^\top  \rg=\lg\calQ_\bot \mZ  \mU^\top  , \mZ  \mU^\top  \rg=\lg { \mZ^\top }\calQ_\bot  \mZ ,\mU^\top   \mU\rg$.   \ding{174} uses  Lemma \ref{nvx:lem:1} by noting that $\mU^\top  \calQ  \mZ  = (\calQ \mU)^\top    \mZ=\mU^\top    \mZ \succeq 0$  satisfying the   assumptions of Lemma \ref{nvx:lem:1}. \ding{175} uses the fact that
		$\|\mU\mU^\top  -(\calQ  \mZ) (\calQ  \mZ)^\top  \|_F^2=\|\mU\mU^\top  -\calQ \mZ\mZ^\top  \calQ\|_F^2\leq \|\mU\mU^\top  -\calQ \mZ\mZ^\top  \calQ\|_F^2+\|\calQ \mZ\mZ^\top  \calQ_\bot\|_F^2=\|\mU\mU^\top  -\calQ \mZ\mZ^\top  \calQ-\calQ \mZ\mZ^\top  \calQ_\bot\|_F^2=\|\mU\mU^\top  -\calQ  \mZ   \mZ^\top  \|_F^2$. \ding{176} uses the following basic inequality that
		\[\frac{1}{8}\|\mA\|_F^2 +2 \|\mB\|_F^2 \geq 2\sqrt{\frac{2}{8}\|\mA\|_F^2\|\mB\|_F^2}=\|\mA\|_F\|\mB\|_F\geq\lg \mA,\mB\rg,\]
		where $\mA= { \mZ^\top }\calQ_\bot  \mZ$ and $\mB=\mU^\top   \mU-{ \mZ^\top }\calQ \mZ.$

		\paragraph{The Remaining Steps.}
		The remaining steps involve   showing the following bounds:
		\begin{align}
		\|{ \mZ^\top }\calQ_\bot  \mZ \|_F^2&\leq\|\mU \mU^\top  -\mZ  { \mZ^\top }  \|_F^2, \label{nvx:eqn:first:bound}\\
		\lg { \mZ^\top }\calQ_\bot  \mZ ,{ \mZ^\top }\calQ  \mZ \rg&\leq \|\mU \mU^\top  -\calQ \mZ  { \mZ^\top } \|_F^2,\label{nvx:eqn:second:bound}\\
		\|\mU^\top   \mU-{ \mZ^\top }\calQ  \mZ \|_F^2&\leq\|\mU\mU^\top  -\calQ  \mZ  \mZ^\top  \|_F^2.\label{nvx:eqn:third:bound}
		\end{align}
		This is because when plugging these bounds~\eqref{nvx:eqn:first:bound}-\eqref{nvx:eqn:third:bound} into~\eqref{nvx:eqn:final:step},
		we can obtain the desired result:
		\[
		\|({\mU} -  \mZ ){\mU}^\top  \|_F^2 \leq \frac{1}{8}\|\mU\mU^\top   -   \mZ  \mZ^\top    \|_F^2 + \left(3 + \frac{1}{2\sqrt{2} -2} \right)\|(\mU\mU^\top   -  \mZ  \mZ^\top   ) {Q}{Q}^\top  \|_F^2.
		\]

		\paragraph{Showing~\eqref{nvx:eqn:first:bound}.}
		\begin{align*}
		\|{ \mZ^\top }\calQ_\bot  \mZ \|_F^2&=\lg  \mZ  { \mZ^\top }\calQ_\bot , \calQ_\bot  \mZ  { \mZ^\top }\rg\\
		&\stack{\ding{172}}{=}\lg\calQ_\bot  \mZ  { \mZ^\top }\calQ_\bot , \calQ_\bot  \mZ  { \mZ^\top }\calQ_\bot\rg\\
		&=\|\calQ_\bot  \mZ  { \mZ^\top }\calQ_\bot\|_F^2\\
		&\stack{\ding{173}}{=}\|\calQ_\bot (\mZ  { \mZ^\top }-\mU \mU^\top  )\calQ_\bot\|_F^2\\
		&\stack{\ding{174}}{\leq}\| \mZ  { \mZ^\top }-\mU \mU^\top   \|_F^2,
		\end{align*}
		where \ding{172} follows from the idempotence property that $\calQ_\bot=\calQ_\bot\calQ_\bot.$ \ding{173}
		follows from $\calQ_\bot \mU=\zero$.
		\ding{174} follows from the nonexpansiveness of projection operator: $\|\calQ_\bot (\mZ  { \mZ^\top }-\mU \mU^\top  )\calQ_\bot\|_F\leq \|(\mZ  { \mZ^\top }-\mU \mU^\top  )\calQ_\bot\|_F\leq \|\mZ  { \mZ^\top }-\mU \mU^\top  \|_F$.
		
		\paragraph{Showing~\eqref{nvx:eqn:second:bound}.}  The argument here is pretty similar to that for~\eqref{nvx:eqn:first:bound}:
		\begin{align*}
		\lg { \mZ^\top }\calQ_\bot  \mZ ,{ \mZ^\top }\calQ  \mZ \rg
		&=\lg \calQ  \mZ  { \mZ^\top } ,  \mZ  { \mZ^\top }\calQ_\bot\rg\\
		&=\lg \calQ  \mZ  { \mZ^\top } \calQ_\bot,  \calQ \mZ  { \mZ^\top }\calQ_\bot\rg\\
		&=\|\calQ  \mZ  { \mZ^\top }\calQ_\bot\|_F^2\\
		&\stack{\ding{172}}{=}\|\calQ (\mZ  { \mZ^\top }-\mU \mU^\top  )\calQ_\bot\|_F^2\\
		&\stack{\ding{173}}{\leq} \|\calQ  \mZ  { \mZ^\top }-\mU \mU^\top   \|_F^2,
		\end{align*}
		where
		\ding{172} is by $\calQ_\bot \mU=\zero$.
		\ding{173} uses the  nonexpansiveness of projection operator and $\calQ \mU\mU^\top =\mU\mU^\top .$

		\paragraph{Showing ~\eqref{nvx:eqn:third:bound}.}
		First by expanding $\|\cdot\|_F^2$ using inner products,~\eqref{nvx:eqn:third:bound} is equivalent to the following inequality
		\begin{align}\label{nvx:eqn:third:bound:1}
		\|\mU^\top   \mU\|_F^2+\|\mU^\top   \mU-\mZ^\top \calQ  \mZ\|_F^2-2 \lg \mU^\top   \mU,\mZ^\top \calQ  \mZ\rg \leq \|\mU\mU^\top \|_F^2 +\|\calQ  \mZ  \mZ^\top  \|_F^2-2\lg \mU\mU^\top ,\calQ  \mZ  \mZ^\top \rg.
		\end{align}
		First of all, we recognize that
		\begin{align*}
		&\|\mU^\top  \mU\|_F^2=\sum_i \sigma_i(\mU)^2=\|\mU\mU^\top \|_F^2;\\
		&\|{ \mZ^\top }\calQ \mZ \|_F^2=\lg { \mZ^\top }\calQ \mZ,{ \mZ^\top }\calQ \mZ\rg=\lg  \calQ \mZ\mZ^\top  , \mZ{ \mZ^\top }\calQ  \rg=\lg  \calQ \mZ\mZ^\top  \calQ, \mQ \mZ{ \mZ^\top }\calQ  \rg=\|\calQ \mZ   \mZ^\top \calQ\|_F^2\leq\| \mZ  \mZ^\top \calQ\|_F^2,
		\end{align*}
		where we use the idempotence and nonexpansiveness property of the projection matrix $\calQ$ in the second line.
		Plugging these to~\eqref{nvx:eqn:third:bound:1}, we find ~\eqref{nvx:eqn:third:bound:1} reduces to
		\begin{align}\label{nvx:eqn:third:bound:2}
		\lg \mU^\top   \mU, { \mZ^\top }\calQ  \mZ \rg\geq \lg \mU\mU^\top  , \calQ  \mZ  { \mZ^\top }\rg=\lg \mU\mU^\top  ,  \mZ  { \mZ^\top }\rg= \|\mU^\top  \mZ\|_F^2.
		\end{align}
		To show~\eqref{nvx:eqn:third:bound:2}, let $\mQ\mSigma \mP^\top $ be the SVD of $\mU$ with $\mSigma\in\R^{r' \times r'}$ and $\mP\in\R^{r\times r'}$ where $r'$ is rank of $\mU$.  Then
		\begin{align}\label{nvx:eqn:third:bound:QW}
		\mU^\top  \mU=\mP\mSigma^2\mP^\top  , \qquad   \mQ=\mU\mP\mSigma\inv\qquad  \text{and} \qquad\calQ=\mQ\mQ^\top =\mU\mP\mSigma^{-2}\mP^\top  \mU^\top  .
		\end{align}
		Now
		\begin{align*}
		\tmop{LHS~of~}\eqref{nvx:eqn:third:bound:2}
		&= \lg \mU^\top   \mU, { \mZ^\top }\calQ  \mZ \rg\\
		&\stack{\ding{172}}{=}\lg \mP\mSigma^2\mP^\top  , { \mZ^\top }UP\mSigma^{\text{-}2}\mP^\top  \mU^\top  \mZ \rg\\
		&\stack{\ding{173}}{=}\lg \mSigma^2, \mP^\top  (\mU^\top   \mZ )P\mSigma^{\text{-}2}\mP^\top  (\mU^\top  \mZ )\mP\rg\\
		&\stack{\ding{174}}{=}\lg \mSigma^2, \mG\mSigma^{\text{-}2}\mG\rg\\
		& =\|\mSigma \mG\mSigma\inv\|_F^2\\
		&\stack{\ding{175}}{\geq}\|\mG\|_F^2\\
		&\stack{\ding{176}}{=}\|\mU^\top   \mZ \|_F^2,
		\end{align*}
    	where
   	\ding{172} is by  ~\eqref{nvx:eqn:third:bound:QW} and
	\ding{173} uses the assumption that ${ \mZ^\top }\mU=\mU^\top   \mZ\succeq 0.$
	In \ding{174}, we define $\mG:=\mP^\top  (\mU^\top   \mZ )\mP$.
	\ding{176} is because $\|\mG\|_F^2=\|\mP^\top  (\mU^\top   \mZ )\mP\|_F^2=\|\mU^\top  \mZ \|_F^2$ due to the rotational invariance of $\|\cdot\|_F.$
	\ding{175} is because
		\begin{align*}
		\|\mSigma \mG\mSigma\inv\|_F^2&=\sum_{i,j}\frac{\sigma_i^2}{\sigma_j^2} G_{ij}^2\\
		&=\sum_{i=j}G_{ii}^2+\sum_{i> j}\left( \frac{\sigma_i^2}{\sigma_j^2} +\frac{\sigma_j^2}{\sigma_i^2} \right)G_{ij}^2\\
		&\geq \sum_{i=j}G_{ii}^2+\sum_{i> j}2\left( \frac{\sigma_i}{\sigma_j} \right)\left( \frac{\sigma_j}{\sigma_i} \right)G_{ij}^2\\
		&=\sum_{i,j}G_{ij}^2\\
		&=\|\mG\|_F^2,
		\end{align*}
		where the second line follows from the symmetric property of $\mG$ since $\mG=\mP^\top  (\mU^\top   \mZ )\mP\succeq 0$ and $\mU^\top   \mZ \succeq 0$.
	\end{proof}

	\section{Proof of Lemma \ref{nvx:lem:3}}\label{nvx:sec:proof:lem3}
	
	Let $\mX=\mU\mU^\top $ and $\mX^\star= {\mU}^\star \mU^{\star \top}.$   We start with the critical point  condition $\nabla f(\mX)\mU=\zero$ which implies
	\[\nabla f(\mX)\mU\mU^{\dagger}=\nabla f(\mX)\mQ\mQ^\top =\zero,\]
	where $^{\dagger}$ denotes the pseudoinverse.   Then for all $Z\in\R^{n\times n}$, we have
    {\allowdisplaybreaks
	\begin{align*}
	&\Rightarrow \lg \nabla f(\mX),\mZ \mQ\mQ^\top  \rg=0\\
	&\stack{\ding{172}}{\Rightarrow} \lg\nabla f(\mX^\star)+\int_0^1 [\nabla^2f(t \mX + (1-t)\mX^\star)](\mX-\mX^\star)d t,\mZ \mQ\mQ^\top  \rg=0\\
	&\Rightarrow  \lg\nabla f(\mX^\star),\mZ \mQ\mQ^\top  \rg+ \left[\int_0^1\nabla^2 f(t \mX + (1-t)\mX^\star)d t\right](\mX-\mX^\star,\mZ \mQ\mQ^\top  )=0\\
	&\stack{\ding{173}}{\Rightarrow} \left| -\frac{2}{\beta+\alpha}\lg\nabla f(\mX^\star),\mZ \mQ\mQ^\top  \rg -\lg \mX-\mX^\star, \mZ\mQ\mQ^\top  \rg \right|\leq \frac{\beta-\alpha}{\beta+\alpha}\|\mX-\mX^\star\|_F\|Z\mQ\mQ^\top  \|_F\\
	&\Rightarrow  \left| \frac{2}{\beta+\alpha}\lg\nabla f(\mX^\star),\mZ \mQ\mQ^\top  \rg +\lg \mX-\mX^\star, \mZ\mQ\mQ^\top  \rg \right|\leq \frac{\beta-\alpha}{\beta+\alpha}\|\mX-\mX^\star\|_F\|\mZ\mQ\mQ^\top  \|_F\\
	&\stack{\ding{174}}{\Rightarrow}   \left|  \frac{2}{\beta+\alpha}\lg\nabla f(\mX^\star),(\mX-\mX^\star)\mQ\mQ^\top  \rg+\|(\mX-\mX^\star)\mQ\mQ^\top  \|_F^2  \right|\leq  \frac{\beta-\alpha}{\beta+\alpha}\|\mX-\mX^\star\|_F \|(\mX-\mX^\star)\mQ\mQ^\top  \|_F\\
	&\stack{\ding{175}}{\Rightarrow}    \frac{2}{\beta+\alpha}\lg\nabla f(\mX^\star),(\mX-\mX^\star)\mQ\mQ^\top  \rg+\|(\mX-\mX^\star)\mQ\mQ^\top  \|_F^2  \leq  \frac{\beta-\alpha}{\beta+\alpha}\|\mX-\mX^\star\|_F \|(\mX-\mX^\star)\mQ\mQ^\top  \|_F
	\\
	&\Rightarrow  \|(\mX-\mX^\star)\mQ\mQ^\top  \|_F  \leq  \delta \|\mX-\mX^\star\|_F,
	\end{align*}
    }
	where
	\ding{172}  uses the Taylor's Theorem for vector-valued functions ~\cite[Eq. (2.5) in Theorem 2.1]{nvx:nocedal2006numerical}.
	\ding{173}  uses Proposition \ref{nvx:pro:RIP} by noting that the PSD matrix $[t \mX^\star+ (1-t)\mX]$ has rank at most $2r$ for all $t\in[0,1]$ and  $\rank(\mX-\mX^\star)\leq 4r,\rank(\mZ\mQ\mQ^\top )\leq 4r$.
	\ding{174}  is by choosing $\mZ=X-\mX^\star.$
	\ding{175}  follows from $\lg\nabla f(\mX^\star),(\mX-\mX^\star)\mQ\mQ^\top  \rg\geq0$ since
	\begin{align*}
	\lg \nabla f(\mX^\star), (\mX-\mX^\star)\mQ\mQ^\top  \rg\stack{(i)}{=}\lg \nabla f(\mX^\star), \mX- \mX^\star \mQ\mQ^\top  \rg\stack{(ii)}{=}\lg \nabla f(\mX^\star), \mX\rg \stack{(iii)}{\geq}0,
	\end{align*}
	where
	(i) follows from $X\mQ\mQ^\top =\mU\mU^\top \mQ\mQ^\top =\mU\mU^\top $ since $\mQ\mQ^\top $ is the orthogonal projector onto  $\range(\mU)$.
	(ii) uses the fact that
	\[\nabla f(\mX^\star) \mX^\star=\zero=\mX^\star \nabla f(\mX^\star),\]
	and
	(iii) is because $\nabla f(\mX^\star)\succeq 0, \mX\succeq 0$.
	\hfill$\square$

	\section{Proof of Proposition \ref{nvx:pro:ef}} \label{nvx:sec:proof:pro:ef}
	For any critical point $(\mU,\mV)$, we have
	$$\nabla g(\mU,\mV)=\Xi(\mU\mV^\top )\mW=\zero,$$
	where $\mW=\begin{bmatrix} \mU^\top &\mV^\top  \end{bmatrix}^\top$. Further denote $\wh{\mW}=\begin{bmatrix} \mU^\top &-\mV^\top  \end{bmatrix}^\top$. Then
	\begin{align*}
	\stack{\ding{172}}{\Rightarrow}&\wh{\mW}^\top \nabla g(\mU,\mV)+\nabla g(\mU,\mV)^\top \wh{\mW}=\zero\\
	\stack{\ding{173}}{\Rightarrow}&  \wh{\mW}^\top \Xi(\mU\mV^\top )\mW+\mW^\top \Xi(\mU\mV^\top )\wh{\mW}=\zero  \\
	\stack{\ding{174}}{\Rightarrow}& [\mU^\top ~-\mV^\top ]\begin{bmatrix}
	\lambda\eye&\nabla f(\mU\mV^\top )\\
	\nabla f(\mU\mV^\top )^\top &\lambda\eye
	\end{bmatrix}\begin{bmatrix}\mU\\ \mV\end{bmatrix}+
	[\mU^\top ~\mV^\top ]\begin{bmatrix}
	\lambda\eye&\nabla f(\mU\mV^\top )\\
	\nabla f(\mU\mV^\top )^\top &\lambda\eye
	\end{bmatrix}\begin{bmatrix}\mU\\ -\mV\end{bmatrix}=\zero\\
	\stack{\ding{175}}{\Rightarrow}& \lambda\left(2\mU^\top \mU-2\mV^\top \mV\right)+\underbrace{\mU^\top \left(\nabla f(\mU\mV^\top )-\nabla f(\mU\mV^\top )\right)\mV}_{=\zero}
	+\underbrace{\mV^\top \left(\nabla f(\mU\mV^\top )^\top -\nabla f(\mU\mV^\top )^\top \right)\mU}_{=\zero}=\zero\\
	{\Rightarrow}& 2\lambda(\mU^\top \mU-\mV^\top \mV)=\zero\\
	\stack{\ding{176}}{\Rightarrow}& \mU^\top \mU-\mV^\top \mV =\zero,
	\end{align*}
	where \ding{172} follows from $\nabla g(\mU,\mV)=\zero$ and
	\ding{173} follows from $\nabla g(\mU,\mV)=\Xi(\mU\mV^\top )W$.   \ding{174} follows by plugging the definitions of $W,\wh{\mW}$ and $\Xi(\cdot)$ into the second line.
	\ding{175} follows from direct computations. \ding{176} holds since   $\lambda>0.$
	\hfill$\square$

	\section{Proof of Lemma \ref{nvx:lem:ef:1}} \label{nvx:sec:proof:lem:ef:1}
	
	First recall
	$$
	{\mW}=\begin{bmatrix}\mU\\ \mV \end{bmatrix},\qquad \wh{\mW}=\begin{bmatrix}\mU\\ -\mV \end{bmatrix},\qquad {\mD}=\begin{bmatrix}\mD_\mU\\\mD_\mV \end{bmatrix},\qquad \wh{\mD}=\begin{bmatrix}\mD_\mU\\ -\mD_\mV \end{bmatrix}.
	$$
	By performing the following change of variables
	\begin{align*}
		\mW_1&\leftarrow \mD,\qquad
	\wh{\mW}_1\leftarrow \wh{\mD},\qquad
	\mW_2\leftarrow	\mW,\qquad
	\wh{\mW}_2\leftarrow\wh{\mW}
	\end{align*}
	in~\eqref{nvx:eqn:Pon:Poff}, we have
	\begin{align*}
	\|\Pon(\mD\mW^\top )\|_F^2&=\frac{1}{4}\| \mD\mW^\top +\wh{\mD}\wh{\mW}^\top \|_F^2 =\frac{1}{4}\lg \mD\mW^\top +\wh{\mD}\wh{\mW}^\top , \mD\mW^\top +\wh{\mD}\wh{\mW}^\top  \rg;\\
	\|\Poff(\mD\mW^\top )\|_F^2&=\frac{1}{4}\| \mD\mW^\top -\wh{\mD}\wh{\mW}^\top \|_F^2=\frac{1}{4}\lg \mD\mW^\top -\wh{\mD}\wh{\mW}^\top , \mD\mW^\top -\wh{\mD}\wh{\mW}^\top  \rg.
	\end{align*}
	Then it implies that
	\begin{align*}
	\|\Pon(\mD\mW^\top )\|_F^2-\|\Poff(\mD\mW^\top )\|_F^2
	&=\frac{1}{4}\lg \mD\mW^\top +\wh{\mD}\wh{\mW}^\top , \mD\mW^\top +\wh{\mD}\wh{\mW}^\top  \rg-\frac{1}{4}\lg \mD\mW^\top -\wh{\mD}\wh{\mW}^\top , \mD\mW^\top -\wh{\mD}\wh{\mW}^\top  \rg\\
	&= \lg \mD\mW^\top , \wh{\mD}\wh{\mW}^\top  \rg
	= \lg \wh{\mD}^\top \mD , \wh{\mW}^\top \mW \rg
	=0,
	\end{align*}
   since $\wh{\mW}^\top W =\zero$   from~\eqref{nvx:eqn:balanced:set}.
	\hfill$\square$

	\section{Proof of Lemma \ref{nvx:lem:ef:2}} \label{nvx:sec:proof:lem:ef:2}
	To begin with, we define $\wh{\mW}_1=\begin{bmatrix}\mU_1\\ -\mV_1 \end{bmatrix}$, $\wh{\mW}_2=\begin{bmatrix}\mU_2\\ -\mV_2 \end{bmatrix}$. Then
	\begin{align*}
	&\|\Pon(\mW_1\mW_1^\top -\mW_2 \mW_2^\top )\|_F^2-\|\Poff(\mW_1\mW_1^\top -\mW_2 \mW_2^\top )\|_F^2\\
	&\stack{\ding{172}}{=}\|\Pon(\mW_1\mW_1^\top )-\Pon(\mW_2 \mW_2^\top )\|_F^2-\|\Poff(\mW_1\mW_1^\top )-\Poff(\mW_2 \mW_2^\top )\|_F^2\\
	&\stack{\ding{173}}{=}\left\|\frac{\mW_1\mW_1^\top +\wh{\mW}_1\wh{\mW}_1^\top }{2}-\frac{\mW_2\mW_2^\top +\wh{\mW}_2\wh{\mW}_2^\top }{2}\right\|_F^2-
	\left\|\frac{\mW_1\mW_1^\top -\wh{\mW}_1\wh{\mW}_1^\top }{2}-\frac{\mW_2\mW_2^\top -\wh{\mW}_2\wh{\mW}_2^\top }{2}\right\|_F^2\\
	&=\left\|\frac{\mW_1\mW_1^\top -\mW_2\mW_2^\top }{2}+\frac{\wh{\mW}_1\wh{\mW}_1^\top -\wh{\mW}_2\wh{\mW}_2^\top }{2}\right\|_F^2-
	\left\|\frac{\mW_1\mW_1^\top -\mW_2\mW_2^\top }{2}-\frac{\wh{\mW}_1\wh{\mW}_1^\top -\wh{\mW}_2\wh{\mW}_2^\top }{2}\right\|_F^2\\
	&\stack{\ding{174}}{=}\lg \mW_1\mW_1^\top -\mW_2 \mW_2^\top ,\wh{\mW}_1\wh{\mW}_1^\top -\wh{\mW}_2 \wh{\mW}_2^\top \rg\\
	&=\lg \mW_1\mW_1^\top ,\wh{\mW}_1\wh{\mW}_1^\top \rg+\lg \mW_2  \mW_2^\top ,\wh{\mW}_2  \wh{\mW}_2^\top \rg
	-\lg \mW_1\mW_1^\top ,\wh{\mW}_2 \wh{\mW}_2^\top \rg-\lg \wh{\mW}_1\wh{\mW}_1^\top	,\mW_2  \mW_2^\top \rg\\
	&\stack{\ding{175}}{=}-\lg \mW_1\mW_1^\top ,\wh{\mW}_2 \wh{\mW}_2^\top \rg-\lg \wh{\mW}_1\wh{\mW}_1^\top	,\mW_2  \mW_2^\top \rg\\
	&\stack{\ding{176}}{\leq} 0,
	\end{align*}
	where \ding{172} is due to the linearity of $\Pon$ and $\Poff$. \ding{173} follows from~\eqref{nvx:eqn:Pon:Poff}.  \ding{174} is by expanding  $\|\cdot\|_F^2$. \ding{175} comes from
	\eqref{nvx:eqn:balanced:set} that
	\[\wh{\mW}_i^\top \mW_i=\mW^\top _i\wh{\mW}_i=\zero,\qquad \text{ for $i=1, 2$}.\]
	\ding{176} uses the fact that
	\[\mW_1\mW_1^\top \succeq0,\qquad  \wh{\mW}_1\wh{\mW}_1^\top \succeq0,\qquad  \mW_2\mW_2^\top \succeq0,\qquad  \wh{\mW}_2\wh{\mW}_2^\top \succeq0.\]
	
	\hfill$\square$
	
	\section{Proof of Proposition \ref{nvx:pro:global}} \label{nvx:sec:proof:pro:global}
	From~\eqref{nvx:eqn:global:factor}, we have
	\begin{align*}
	\frac{1}{2}\left(\|\mU^\star\|_F^2+\|\mV^\star\|_F^2\right)
	&\stack{\ding{172}}{=}\frac{1}{2}\left(\left\|\mP^\star[\sqrt{\mSigma^\star}~\zero_{r^\star\times(r-r^\star)}] \mR\right\|_F^2+\left\|\mQ^\star[\sqrt{\mSigma^\star}~\zero_{r^\star\times(r-r^\star)}] \mR\right\|_F^2\right)\\
	&\stack{\ding{173}}{=}\frac{1}{2}\left(\left\|\sqrt{\mSigma^\star}\right\|_F^2+\left\|\sqrt{\mSigma^\star}\right\|_F^2\right)\\
	&=\left\|\sqrt{\mSigma^\star}\right\|_F^2\\
	&\stack{\ding{174}}{=}\|\mX^\star\|_*,
	\end{align*}
	where \ding{172} uses the definitions of $\mU^\star$ and $\mV^\star$ in~\eqref{nvx:eqn:global:factor}. \ding{173} uses the rotational invariance of $\|\cdot\|_F.$ \ding{174} is because $\|\sqrt{\mSigma^\star}\|_F^2=\sum_j \sigma_k(\mX^\star)=\|\mX^\star\|_*.$
	
	Therefore,
	\begin{align*}
	f(\mU^\star \mV^{\star \top})+\lambda (\|\mU^\star\|_F^2+\|V^\star\|_F^2)/2
	&\stack{\ding{172}}{=}f(\mX^\star)+\lambda\|\mX^\star\|_*\\
	&\leq f(\mX)+\lambda\|\mX\|_*\\
	&\stack{\ding{173}}{=} f(\mU\mV^\top )+\lambda\|\mU\mV^\top \|_*\\
	&\stack{\ding{174}}{\leq} f(\mU\mV^\top )+\lambda(\|\mU\|_F^2+\|\mV\|_F^2)/2,
	\end{align*}
	where \ding{172} comes from the optimality of $\mX^\star$ for~\eqref{nvx:eqn:P1}. \ding{173} is by choosing $\mX=\mU\mV^\top .$  \ding{174} is because  $\|\mU\mV^\top \|_*\leq (\|\mU\|_F^2+\|\mV\|_F^2)/2$ by the optimization formulation of the matrix nuclear norm ~\cite[Lemma 5.1]{nvx:recht2010guaranteed} that
	\[\|\mX\|_*=\min_{\mX=\mU\mV^\top } \frac{1}{2}(\|\mU\|_F^2+\|\mV\|_F^2).\]
	
	\hfill$\square$

	\section{Proof of Lemma \ref{nvx:lem:bound:QQ}}\label{nvx:app:lem:bound:QQ}
	Let $\mZ=\begin{bmatrix}\mZ_\mU\\ \mZ_\mV \end{bmatrix}$ with arbitrary $\mZ_\mU\in\R^{n\times r}$ and $\mZ_\mV\in\R^{m\times r}$. Then

	{\allowdisplaybreaks
	\begin{align*}
	&\Rightarrow
	\lg\Xi(\mX)
	\mW,\mZ\rg=\lg\zero,\mZ\rg=0
	\\
	&\Rightarrow
	\left\lg
	\Xi(\mX)-\Xi(\mX^\star)
	+
	\Xi(\mX^\star)
	,
	\mZ\mW^\top
	\right\rg
	=0
	\\
	&\Rightarrow
	\left\lg
	\begin{bmatrix}
	\lambda\eye&\nabla f(\mX)\\
	\nabla f(\mX)^\top &\lambda\eye
	\end{bmatrix}-
	\begin{bmatrix}
	\lambda\eye&\nabla f(\mX^\star)\\
	\nabla f(\mX^\star)^\top &\lambda\eye
	\end{bmatrix}
	+
	\Xi(\mX^\star)
	,
	\mZ\mW^\top
	\right\rg
	=0
	\\
	&\Rightarrow
	\left\lg
	\begin{bmatrix}
	\zero&\nabla f(\mX)-\nabla f(\mX^\star)\\
	\nabla f(\mX)^\top -\nabla f(\mX^\star)^\top &\zero
	\end{bmatrix}
	+
	\Xi(\mX^\star)
	,
	\mZ\mW^\top
	\right\rg
	=0
	\\
	&\Rightarrow
	\left\lg
	\begin{bmatrix}
	\zero&\int_{0}^1[\nabla^2 f(\mX^\star+t(\mX-\mX^\star))](\mX-\mX^\star)d t\\
	* &\zero
	\end{bmatrix}
	+
	\Xi(\mX^\star)
	,
	\mZ\mW^\top
	\right\rg
	=0
	\\
	&\Rightarrow
	\left\lg
	\begin{bmatrix}
	\zero&\int_{0}^1[\nabla^2 f(\mX^\star+t(\mX-\mX^\star))](\mX-\mX^\star)d t\\
	* &\zero
	\end{bmatrix}
	,
	\begin{bmatrix}
	\mZ_\mU\mU^\top &\mZ_\mU\mV^\top \\
	\mZ_\mV\mU^\top &\mZ_\mV\mV^\top
	\end{bmatrix}
	\right\rg
	+
	\left\lg
	\Xi(\mX^\star)
	,
	\mZ\mW^\top
	\right\rg
	=0
	\\
	&\Rightarrow
	\int_{0}^1[\nabla^2 f(\mX^\star+t(\mX-\mX^\star))](\mX-\mX^\star,\mZ_\mU\mV^\top +\mU\mZ_\mV^\top )d t
	+
	\left\lg
	\Xi(\mX^\star)
	,
	\mZ\mW^\top
	\right\rg
	=0,
	\end{align*}
    }
	where  the fifth line follows from the   Taylor's Theorem for vector-valued functions ~\cite[Eq. (2.5) in Theorem 2.1]{nvx:nocedal2006numerical} and for convenience $* = \left(\int_{0}^1[\nabla^2 f(\mX^\star+t(\mX-\mX^\star))](\mX-\mX^\star)d t\right)^\top $ in the fifth and sixth lines.
	Then, from Proposition \ref{nvx:pro:RIP} and Eq.~\eqref{nvx:eqn:Pon:Poff}, we have
	\begin{equation}\label{nvx:eqn:A:1}
	\begin{aligned}
	&\bigg|\frac{2}{\beta+\alpha}\underbrace{\left\lg
		\Xi(\mX^\star)
		,
		\mZ\mW^\top
		\right\rg}_{\Pi_1(\mZ)}+ \underbrace{\lg\Poff(\mW\mW^\top -\mW^\star \mW^{\star \top}),\mZ\mW^\top \rg}_{\Pi_2(\mZ)}\bigg|
	\leq \frac{\beta-\alpha}{\beta+\alpha}
	\|\mX-\mX^\star\|_F\underbrace{\|\Poff(\mZ\mW^\top )\|_F}_{\Pi_3(\mZ)}.
	\end{aligned}
	\end{equation}

	\paragraph{The Remaining Steps.}
	The remaining steps are choosing $\mZ=(\mW\mW^\top  - \mW^\star \mW^{\star \top}){\mW^\top }^{\dagger}$ and showing the following
	\begin{align}
	\Pi_1(\mZ)&\geq0 \label{nvx:eqn:Pi:1},\\
	\Pi_2(\mZ)&\geq\frac{1}{2}\|(\mW\mW^\top -\mW^\star \mW^{\star \top})\mQ\mQ^\top \|_F^2\label{nvx:eqn:Pi:2},\\
	\Pi_3(\mZ)& \leq \|(\mW\mW^\top -\mW^\star \mW^{\star \top})\mQ\mQ^\top \|_F\label{nvx:eqn:Pi:3}.
	\end{align}
	Then plugging~\eqref{nvx:eqn:Pi:1}-\eqref{nvx:eqn:Pi:3} into~\eqref{nvx:eqn:A:1} yields the desired result:
	\begin{align*}
	\frac{1}{2}\left\|(\mW\mW^\top -\mW^\star \mW^{\star \top})\mQ\mQ^\top \right\|_F^2
	&\leq \frac{\beta-\alpha}{\beta+\alpha}
	\|\mX-\mX^\star\|_F\left\|(\mW\mW^\top -\mW^\star \mW^{\star \top})\mQ\mQ^\top \right\|_F,
	\end{align*}
	or equivalently,
	\[ \left\|(\mW\mW^\top -\mW^\star \mW^{\star \top})\mQ\mQ^\top \right\|_F
	\leq 2\frac{\beta-\alpha}{\beta+\alpha}
	\|\mX-\mX^\star\|_F.\]
	
	\paragraph{Showing~\eqref{nvx:eqn:Pi:1}.}
	Choosing $\mZ=(\mW\mW^\top  - \mW^\star \mW^{\star \top}){\mW^\top }^{\dagger}$ and noting that $\mQ\mQ^\top =\mW^{ T}{\mW^\top }^{\dagger}$, we have $\mZ\mW^\top =(\mW\mW^\top -\mW^\star \mW^{\star \top}){\mW^\top }^{\dagger} \mW^\top =(\mW\mW^\top -\mW^\star \mW^{\star \top})\mQ\mQ^\top $. Then
	\[\Pi_1(\mZ)=\lg\Xi(\mX^\star),(\mW\mW^\top -\mW^\star \mW^{\star \top})\mQ\mQ^\top \rg=\lg\Xi(\mX^\star),\mW\mW^\top \rg\geq0,\]
	where the second equality holds since $\mW\mW^\top \mQ\mQ^\top =\mW\mW^\top $ and $\Xi(\mX^\star) \mW^\star=\zero$ by~\eqref{nvx:eqn:optimality:origin:3}. The inequality is due to $\Xi(\mX^\star)\succeq0$.

	\paragraph{Showing~\eqref{nvx:eqn:Pi:2}.}
	First recognize that $\Poff(\mW\mW^\top -\mW^\star \mW^{\star \top})=\frac{1}{2}( \mW\mW^\top -\mW^\star \mW^{\star \top}-\wh{\mW}\wh{\mW}^\top +\wh{\mW}^\star \wh{\mW}^{\star \top}).$ Then
	\begin{align*}
	\Pi_2(\mZ)&=\lg\Poff(\mW\mW^\top -\mW^\star \mW^{\star \top}),\mZ\mW^\top \rg\\
	&= \frac{1}{2}\left\lg \mW\mW^\top -\mW^\star \mW^{\star \top}, (\mW\mW^\top -\mW^\star \mW^{\star \top})\mQ\mQ^\top \right\rg
	- \frac{1}{2}\left\lg \wh{\mW}\wh{\mW}^\top -\wh{\mW}^\star \wh{\mW}^{\star \top}, (\mW\mW^\top -\mW^\star \mW^{\star \top})\mQ\mQ^\top \right\rg.
	\end{align*}
	Therefore, ~\eqref{nvx:eqn:Pi:2} follows from
	\begin{align*}
	&\left\lg \wh{\mW}\wh{\mW}^\top -\wh{\mW}^\star \wh{\mW}^{\star \top}, (\mW\mW^\top -\mW^\star \mW^{\star \top})\mQ\mQ^\top \right\rg
	=\left\lg \wh{\mW}\wh{\mW}^\top ,-\mW^\star \mW^{\star \top}\right\rg+\left\lg -\wh{\mW}^\star \wh{\mW}^{\star \top},\mW\mW^\top \right\rg
	\leq0,
	\end{align*}
	where the first equality uses~\eqref{nvx:eqn:balanced:set} and the inequality is because
	$$\wh{\mW}\wh{\mW}^\top \succeq 0,\qquad \mW^\star \mW^{\star \top}\succeq 0,\qquad \wh{\mW}^\star \wh{\mW}^{\star \top}\succeq 0,\qquad \mW\mW^\top \succeq0.$$
	
	\paragraph{Showing~\eqref{nvx:eqn:Pi:3}.}
	Plugging $\mZ=(\mW\mW^\top  - \mW^\star \mW^{\star \top}){\mW^\top }^{\dagger}$ gives
	\[\Pi_3(\mZ)=\|\Poff((\mW\mW^\top -\mW^\star \mW^{\star \top})\mQ\mQ^\top )\|_F,\]
	which is obviously no larger than $\|(\mW\mW^\top -\mW^\star \mW^{\star \top})\mQ\mQ^\top \|_F$ by the definition of the operation $\Poff$.

\hfill$\square$

\end{appendices}

\bibliographystyle{plain}
\bibliography{nonconvex} 
\end{document}